\documentclass[pra, aps, amsmath, amssymb, twocolumn, superscriptaddress,nofootinbib]{revtex4-1}

\usepackage{graphicx}
\usepackage{hyperref}
\usepackage{physics}
\usepackage{color}
\usepackage{booktabs}
\usepackage{enumerate}
\usepackage{bm,ulem}
 \usepackage[caption=false]{subfig}

\newcommand{\ba}{\begin{eqnarray}}
\newcommand{\ea}{\end{eqnarray}}

\begin{document}

\title{Structured quantum collision models: generating coherence with thermal resources}

\author{Stefano Cusumano}
\email[Corresponding author: ]{ste.cusumano@gmail.com}
\affiliation{International Centre for Theory of Quantum Technologies, University of Gda\'nsk, Wita Stwosza 63, 80-308 Gda\'nsk, Poland}
\affiliation{Università degli studi di Napoli ``Federico II'', Dipartimento di Fisica ``E. Pancini'', Via Cupa Cinthia 21, 80126 Napoli, Italy}

\author{Gabriele De Chiara}
\affiliation{Centre for Quantum Materials and Technology, School of Mathematics and Physics,
Queen’s University Belfast, Belfast BT7 1NN, United Kingdom}

\begin{abstract}
Quantum collision models normally consist of a system interacting with a set of ancillary units representing the environment. While these ancillary systems are usually assumed to be either two level systems (TLS) or harmonic oscillators, in this work we move further and represent each ancillary system as a structured system, i.e., a system made out of two or more subsystems. We show how this scenario modifies the kind of master equation that one can obtain for the evolution of the open systems. Moreover, we are able to consider a situation where the ancilla state is thermal yet has some coherence. This allows the generation of coherence in the steady state of the open system and, thanks to the simplicity of the collision model, this allows us to better understand the thermodynamic cost of creating coherence in a system. Specifically, we show that letting the system interact with the coherent degrees of freedom requires a work cost, leading to the natural fulfillment of the first and second law of thermodynamics without the necessity of {\it ad hoc} formulations.
\end{abstract}

\maketitle

%%%%%%%%%%%%%%%%%%%%%%%%%%%%%%%%%%%%%%%%%%%%%%%%%%

\section{Introduction}

Open quantum systems are those systems that do not evolve unitarily, i.e. according to the Schr\"odinger equation, due to the presence of a usually very large external system, dubbed the environment, which interacts with the system influencing its dynamical evolution~\cite{blum,breuer}.

The main problem when studying such systems is the very large number of degrees of freedom of the environment, which make it practically impossible to compute the exact dynamics of the joint system. In order to overcome this problem, several techniques have been developed in order to take into account the effects of the environment, both analytical and numerical. Among the analytical methods we mention master equations methods~\cite{alicki}, stochastic differential equations~\cite{gardiner} and path integrals techniques, while among numerical methods one has the density matrix renormalization group~\cite{schwollock,rossini}, matrix product states~\cite{cirac,orus} and other efficient truncation techniques to efficiently simulate the dynamics of the open system~\cite{tamascelli1,tamascelli2}.

Among these methods, growing attention has been given in recent years to collision models~\cite{coll_mod_rev,review_cusumano,campbell}, which proved to be useful also for efficient quantum simulations~\cite{cattaneo}. In these models one depicts the environment as a collection of smaller units, called {\it ancillae}, which interact piecewise with the system giving rise to an open system dynamics. The main strength of collision models is their versatility in describing many different physical situations by acting only on few parameters of the model, like the state of the ancillae, the interaction time and the interaction Hamiltonian between the ancillae and the system or allowing the ancillae to interact more than once with the system. Acting on these few knobs it has been possible to examine different regimes of open system dynamics, ranging from Markovian~\cite{gisin,scarani,ziman} to non-Markovian dynamics~\cite{ciccarello1,altamirano,ciccarello3,ciccarello4}, and many different physical systems, ranging from optical systems~\cite{cusumano1,cusumano2,ciccarello2} to spin chains~\cite{dechiara2}. Collision models have been used to study different phenomena such as dissipative dynamics, bath engineering, quantum darwinism, the difference between local and global Lindblad equations~\cite{cattaneo2}.

A very important field where collision models have found application is quantum thermodynamics~\cite{strasberg,cusumano3,molitor,rodrigues,barra,piccione}. While in this context one is interested in the influence of quantum effects on work production and heat dissipation, collision models have proved useful in investigating the efficiency of quantum heat engines and the role of quantum coherence in thermodynamic phenomena~\cite{dechiara,dechiara3}.

While in most scenarios the environment is represented as a collection of qubits and independent harmonic oscillators, in recent years some attention has been given to the case of structured environments, that is, environments which are not a simple collection of harmonic oscillators, but rather structured quantum systems, for instance by allowing the harmonic oscillators to have interaction among them or by considering lattice systems as environments~\cite{breuer2,strunz,lukin,keeling,vacchini,manatuly,dag}.
Inspired by this, we introduce a collision model in which the environment is depicted as a collection of ancillae which are intrinsically composite systems, that is, for instance, two or more interacting qubits. The  crucial difference with respect to other works in the literature~\cite{filip,li,purkayastha,purkayastha2}, is the presence of interactions between the components of an ancillary system, while in the cited works such interactions are absent.

These structured ancillae may give rise to driving terms in the master equation, providing a connection with the field of open quantum systems with structured environments. We will also use this model to introduce a situation where, in spite of the fact that the ancillae are in a thermal state, yet they possess coherence in the energy basis. This represents an advancement with respect to previous works, where coherence in the ancillae was assumed a priori, thus hindering the work cost of accessing that coherence and leading to violations of the second law of thermodynamics and ad hoc formulations in order to preserve it.
Another novelty with respect to previous literature is that the generation of steady-state coherence is not engineered through a particular form of the interaction between the system and the bath, but rather it is due to the presence in the interaction Hamiltonian of operators that are not eigenoperators of the free Hamiltonian of the ancillae. This also gives rise explicitly to unitary driving terms in the master equation of the open system, making it clear from where the coherence stems from.

This can be used to generate quantum coherence in the system's steady state, using solely thermal resources. Through this model, we are able to clarify the role of coherence in thermodynamics, highlighting how the interaction of the system with an ancilla with coherence requires work in order to be activated. This makes our work consistent with previous works on steady state coherence generation~\cite{filip2} and coherence broadcasting~\cite{lostaglio,spekkens,faist}, and once more confirms that there is no free lunch.

The manuscript is organized as follows. In Sec.~\ref{sec:collision_model} we introduce our model and derive the most generic master equation for the case where the ancillae are structured instead of being simple TLSs or harmonic oscillators, highlighting the main differences between the two scenarios. Then in Sec.~\ref{sec:examples} we show some features of our model through some examples. In Sec.~\ref{sec:thermodynamics} we use our model to explore the role of coherence in work production and heat dissipation. Finally in Sec.~\ref{sec:conclusions} we draw our conclusions and give some outlook for future works.

%%%%%%%%%%%%%%%%%%%%%%%%%%%%%%%%%%%%%%%%%%%%%%%%%%
\section{Collision models with structured ancillae\label{sec:collision_model}}
%%%%%%%%%%%%%%%%%%%%%%%%%%%%%%%%%%%%%%%%%%%%%%%%%%

In what follows we want to analyze the dynamics of a quantum system $S$ interacting with a structured environment $\mathcal{E}$ by using a collision model.
Here by structured environment we mean an environment composed of ancillae which are composite subsystems. As an example, one can think of each ancilla as being composed by two interacting TLSs or harmonic oscillators.

To represent the collision model, we assume to have an Hamiltonian of the form:
\ba
\hat{H}_{tot}=\hat{H}_S+\hat{H}_{\cal E}+\hat{H}_{S\cal E}
\ea
where the Hamiltonians $\hat{H}_{\cal E}, \hat{H}_{S\cal E}$ are of the form:
\ba
\hat{H}_{\cal E}=\sum_i\hat{H}_{E_i}\quad\hat{H}_{S\cal E}=\sum_i\hat{H}_{SE_i}
\ea
The Hamiltonians $\{\hat{H}_{E_i}\}$ are the free Hamiltonians of the ancillae, while $\hat{H}_{SE_i}$ are the interaction Hamiltonians between the system $S$ and the ancillae. In order to compute the master equation describing the dynamics of $S$, we start our analysis from the free Hamiltonian $\hat{H}_{E_i}$ of the ancillae.
Note that, as we are dealing with structured ancillae, each free Hamiltonian $\hat{H}_{E_i}$ will be in general a many-body operator. Thus, we start our analysis by studying the most generic $\hat{H}_{E_i}$ in the next subsection.

%%%%%%%%%%%%%%%%%%%%%%%%%%%%%%%%%%%%%%%%%%%%%%%%%%%%
\subsection{Structured ancillae}
%%%%%%%%%%%%%%%%%%%%%%%%%%%%%%%%%%%%%%%%%%%%%%%%%%%%

\begin{figure*}[!t]
\centering
\includegraphics[width=\textwidth]{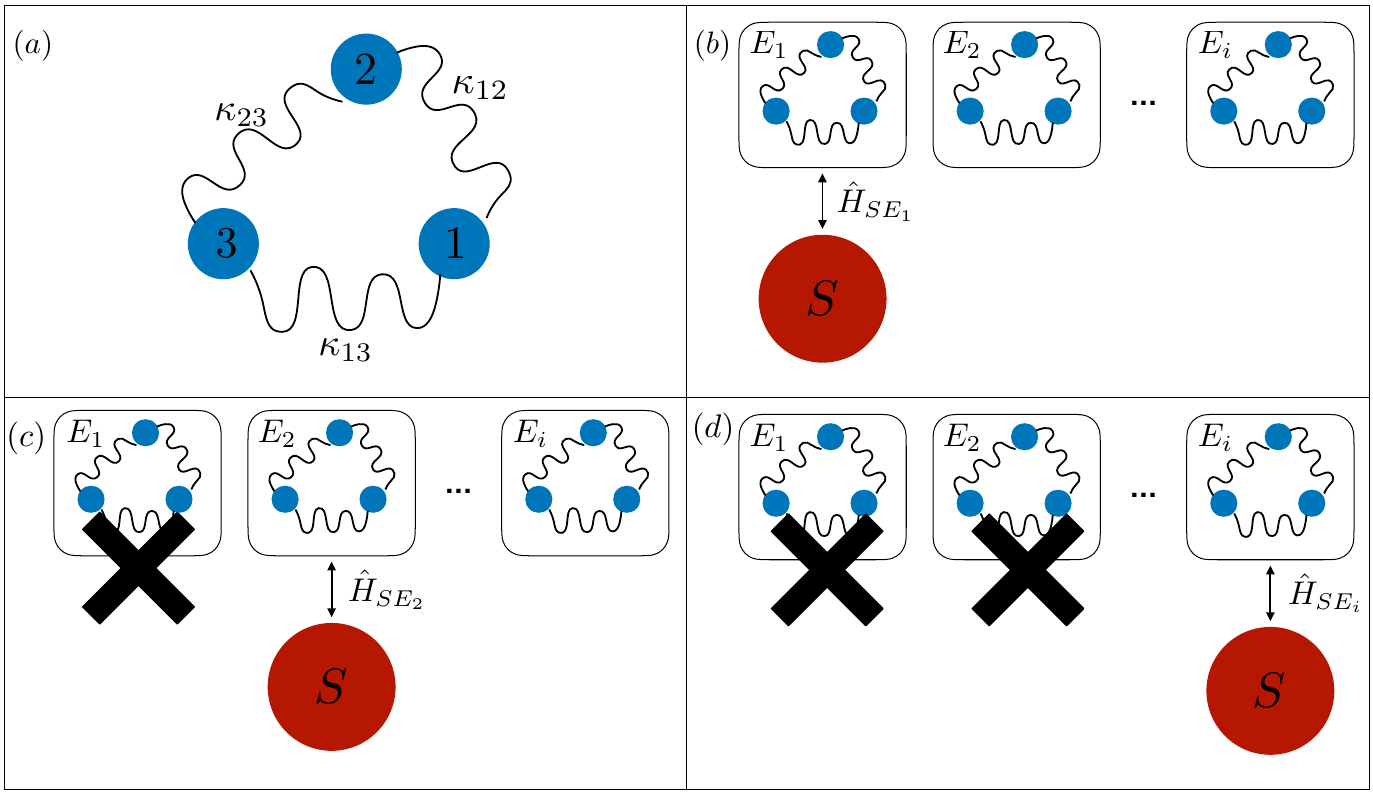}
\caption{In panel (a) one can see an example of structured ancilla made out of three subsystems each interacting with each other. Panels (b)-(d): pictorial representation of the Markovian collision model. In panel (b) the system interacts with the first ancilla, whose state is factorized from the one of all the other ancillas. After the system has interacted with the first ancilla, the latter is traced out and the system interacts with the second ancilla, as in panel (c). The dynamics, first interacting with an ancilla then tracing it out, as in panel (d). }
\label{fig:markovian_composite}
\end{figure*}

In our model, we want to write the Hamiltonian of each ancilla in the form:
\ba
\hat{H}_{E_i}&=&\hat{H}_{E_i}^{\text{free}}+\hat{H}_{E_i}^{\text{int}},\\
\hat{H}_{E_i}^{\text{free}}&=&\sum_{k=1}^K\hbar\omega_k\hat{C}_{E_ik}^\dag\hat{C}_{E_ik},\\
\hat{H}_{E_i}^{\text{int}}&=&\sum_{j\neq k=1}^K\kappa_{jk}\hat{D}_{E_ij}\hat{D}_{E_ik}^\dag+h.c.
\ea
Here the operator $\hat{H}_{E_i}^{\text{free}}$ includes the free Hamiltonians of each subsystem composing the ancilla, while the operator $\hat{H}_{E_i}^{\text{int}}$ contains all the interaction terms between the subsystems of the ancilla.

Note that we are not making any assumption over the nature of the subsystems of the ancillae, as the operators $\hat{C}_k$ can account for both the bosonic and fermionic case by imposing respectively the canonical commutation relations (CCR) or the canonical anti-commutation relations (CAR), that is:
\ba
\nonumber
\comm{\hat{C}_{E_ik}}{\hat{C}_{E_ij}^\dag}=\delta_{jk},\;\;\comm{\hat{C}_{E_ik}}{\hat{C}_{E_ij}}=\comm{\hat{C}_{E_ik}^\dag}{\hat{C}_{E_ij}^\dag}=0,\\
\ea
 for the case of bosons and
\ba
\nonumber
\acomm{\hat{C}_{E_ik}}{\hat{C}_{E_ij}^\dag}=\delta_{jk},\;\;\acomm{\hat{C}_{E_ik}}{\hat{C}_{E_ij}}=\acomm{\hat{C}_{E_ik}^\dag}{\hat{C}_{E_ij}^\dag}=0.\\
\ea
for the case of fermions, where $\comm{\cdot}{\cdot}$ is the commutator between the operators, while $\acomm{\cdot}{\cdot}$ stands for the anti-commutator.

Let us now consider the trivial case of unstructured ancillas, that is to say when $\kappa_{jk}=0\;\forall j,k$. In this case we can easily define the energy eigenbasis of the ancilla as given by product states of each subsystems energy eigenstates:
\ba
\ket{n_1,n_2,\cdots,n_{K}}=\ket{\vec{n}}=\bigotimes_{k=1}^K\ket{n_k}.
\ea
where $\ket{n_k}$ is the n-th energy eigenstate of the k-th subsystem.

However, as soon as interactions are present, i.e. for $\kappa_{jk}\neq0$, the Hamiltonian $\hat{H}_{E_i}$ is not diagonal anymore in the basis $\{\ket{\vec{n}}\}$. This implies that a thermal state of the form $\exp[-\beta\hat{H}_{E_i}]/Z$ is  non-diagonal in the basis $\{\ket{\vec{n}}\}$ as well, so that if the interaction between the ancilla $E_i$ and the system $S$ $\hat{H}_{SE_i}$ is defined in terms of eigenoperators of the Hamiltonian $\hat{H}_{E_i}^{\text{free}}$, a unitary driving term may appear in the master equation for $S$.

To continue setting up the notation, we define as $\{\ket{E'_j}\}$ the eigenbasis of the total Hamiltonian $\hat{H}_{E_i}$, with corresponding eigenvalues $E'_j$.
The eigenbasis of $\hat{H}_{E_i}$ and $\hat{H}_{E_i}^{\text{free}}$ are related via a unitary transformation $\hat{V}^{E_i}$:
\ba
\ket{E'_k}=\sum_{\vec{n}}V^{E_i}_{k\vec{n}}\ket{\vec{n}}.
\ea
Practically, the columns of the matrix $\hat{U}^{E_i}$ contain the coefficients of the expansion of the state $\ket{E'_k}$ in terms of the non-interacting basis $\{\ket{\vec{n}}\}$.

Once we have completed our analysis of the Hamiltonian $\hat{H}_{E_i}$ of the ancillae, we can move to the interaction between $S$ and the ancillae.

%%%%%%%%%%%%%%%%%%%%%%%%%%%%%%%%%%%%%%%%%%%%%%%%
\subsection{System-ancilla interaction and Markovian dynamics}
%%%%%%%%%%%%%%%%%%%%%%%%%%%%%%%%%%%%%%%%%%%%%%%%

We assume a system-ancilla Hamiltonian of the form:
\ba
\label{eq:interaction_hamiltonian}
\hat{H}_{SE_i}=\sum_{j,k}\hat{A}_S^{(k,j)}\otimes\hat{B}_{E_ik}^{(j)},
\ea
where the only assumption is that the Hamiltonian $\hat{H}_{SE_i}$ is Hermitian.

We also assume the initial states of $S$ and the ancillae to be factorized:
\ba
\hat{R}_{S\cal E}(0)=\hat{\rho}_S(0)\bigotimes_{i=1}\hat{\eta}_{E_i}.
\ea
More generally, since after the interaction the ancilla is discarded, we indicate the joint state of $S$ and $\mathcal{E}$ after the collision as:
\ba
\hat{R}_{S\cal E}(i)=\hat{\rho}_S(i)\bigotimes_{j=i+1}\hat{\eta}_{E_j}.
\ea
The joint state of $S$ and $\mathcal{E}$ evolves unitarily according to:
\ba
\hat{R}_{S\cal E}(i)\rightarrow\hat{U}_{SE_{i+1}}\hat{R}_{S\cal E}(i)\hat{U}_{SE_{i+1}}^\dag,
\ea
where the unitary operator $\hat{U}_{SE_i}$ is defined as:
\ba
\label{eq:interaction_unitary}
\hat{U}_{SE_i}=\exp\left[-\frac{i}{\hbar}\delta t\left(\hat{H}_S+\hat{H}_{E_i}+\hat{H}_{SE_{i}}\right)\right].
\ea
%\hat{U}_{SE_i}=\exp\left[-\frac{i}{\hbar}\hat{H}_{SE_i}g\delta t\right].

Here $\delta t$ is the collision time during which $S$ and $E_i$ interact.

In order to connect the state of the system $S$ before and after its interaction with an ancilla, we can expand the unitary operator in Eq.~\eqref{eq:interaction_unitary} in powers of $\delta t$, obtaining:
\ba
\nonumber
\hat{U}_{SE_i}&\simeq&\hat{1}-\frac{i}{\hbar}\delta t(\hat{H}_S+\hat{H}_{E_i}+\hat{H}_{SE_i})\\
\label{eq:unitary_power_expansion}
&&-\frac{\delta t^2}{2\hbar^2}(\hat{H}_S+\hat{H}_{E_i}+\hat{H}_{SE_i})^2,
%\hat{U}_{SE_i}=\hat{1}-\frac{i}{\hbar}g\delta t\hat{H}_{SE_i}-\frac{(g\delta t)^2}{2\hbar^2}\hat{H}_{SE_i}^2+\order{(g\delta t)^3}.\;\;
\ea
where $\hat{1}$ is the identity operator. The dynamics now proceeds in discrete steps. First, the system $S$ interacts with the first ancilla $E_1$. After the interaction the degrees of freedom of $E_1$ are traced away, leaving us with the state of the system after the first collision $\hat{\rho}_S(1)$. Then the system interacts with the second ancilla $E_2$, which is then traced away to leave us with the state $\hat{\rho}_S(2)$ and so on. After the $i$-th collision one has:
\ba
\label{eq:recursive_dynamics}
\hat{\rho}_S(i+1)=\Tr_{E_{i+1}}\left[\hat{U}_{SE_{i+1}}\hat{R}_{S\cal E}(i)\hat{U}_{SE_{i+1}}^\dag\right].
\ea
To obtain a master equation for $\hat{\rho}_S$, we can use the power expansion in Eq.~\eqref{eq:unitary_power_expansion} and substitute it into Eq.~\eqref{eq:recursive_dynamics}. Retaining only terms up to $\delta t^2$ one obtains:
\ba
&&\hat{U}_{SE_{i+1}}\hat{R}_{S\cal E}(i)\hat{U}_{SE_{i+1}}^\dag=\\
\nonumber
&&\mathcal{U}_0\left(\hat{R}_{S\cal E}(i)\right)+\mathcal{U}_1\left(\hat{R}_{S\cal E}(i)\right)+\mathcal{U}_2\left(\hat{R}_{S\cal E}(i)\right),
\ea
where the superoperators $\mathcal{U}_0,\mathcal{U}_1,\mathcal{U}_2$ are the zero, first and second order contributions to the dynamics respectively. The explicit form of these contributions reads:
\ba
&&\mathcal{U}_0\left(\hat{R}_{S\cal E}(i)\right)=\hat{R}_{S\cal E}(i),\\
%&&\mathcal{U}_1\left(\hat{R}_{S\cal E}(i)\right)=-\frac{i}{\hbar}g\delta t\comm{\hat{H}_{SE_{i+1}}}{\hat{R}_{S\cal E}(i)}\\
\nonumber
&&\mathcal{U}_1(\hat{R}_{S\cal E}(i))=\\
\label{eq:schroedinger_fo}
&&\qquad\qquad-\frac{i}{\hbar}\delta t\comm{\hat{H}_S+\hat{H}_{E_{i+1}}+\hat{H}_{SE_{i+1}}}{\hat{R}_{S\cal E}(i)},\\
\nonumber
&&\mathcal{U}_2(\hat{R}_{S\cal E}(i))=\frac{\delta t^2}{2\hbar^2}\\
\nonumber
&&\Big(2(\hat{H}_S+\hat{H}_{E_{i+1}}+\hat{H}_{SE_{i+1}})\hat{R}_{S\cal E}(i)(\hat{H}_S+\hat{H}_{E_{i+1}}+\hat{H}_{SE_{i+1}})\\
&&-\acomm{(\hat{H}_S+\hat{H}_{E_{i+1}}+\hat{H}_{SE_{i+1}})^2}{\hat{R}_{S\cal E}(i)}\Big).\label{eq:schroedinger_so}
%&&\mathcal{U}_2\left(\hat{R}_{S\cal E}(i)\right)=\\
%\nonumber
%&&\frac{(g\delta t)^2}{2\hbar^2}\Bigg(2\hat{H}_{SE_{i+1}}\hat{R}_{S\cal E}(i)\hat{H}_{SE_{i+1}}-\acomm{\hat{H}^2_{SE_{i+1}}}{\hat{R}_{S\cal E}(i)}\Bigg)\;\;.
\ea
It is immediate to observe that the zeroth order contribution corresponds simply to the identity superoperator, while the first order contribution has the form of a unitary contribution to the dynamics. Finally, the second order contribution is the one describing dissipative dynamics.

After computing all these contributions one can trace out the degrees of freedom of $E_{i+1}$, obtaining:
\ba
\nonumber
&&\hat{\rho}_S(i+1)=\\
\label{eq:general_me}
&&\mathcal{L}_0(\hat{\rho}_S(i))+\mathcal{L}_1(\hat{\rho}_S(i))+\mathcal{L}_2(\hat{\rho}_S(i)),
\ea
where once again the superoperators $\mathcal{L}_0,\mathcal{L}_1,\mathcal{L}_2$ represent the zero, first and second order contribution to the dynamics respectively.
Using the expression of the interaction Hamiltonian in Eq.~\eqref{eq:interaction_hamiltonian}, we can write explicitly these contributions as:
\ba
&&\mathcal{L}_0(\hat{\rho}_S(i))=\hat{\rho}_S(i),\\
%&&\mathcal{L}_1(\hat{\rho}_S(i))=-\frac{i}{\hbar}\sum_{k,j}g_{k}^{(j)}\comm{\hat{A}_S^{(k,j)}}{\hat{\rho}_S(i)}\\
\label{eq:schroedinger_first_order}
&&\mathcal{L}_1(\hat{\rho}_S(i))=-\frac{i}{\hbar}\delta t\comm{\hat{H}_S+\sum_{k,j}g_k^{(j)}\hat{A}_S^{(k,j)}}{\hat{\rho}_S(i)},\\
\nonumber
&&\mathcal{L}_2(\hat{\rho}_S(i))=\frac{\delta t^2}{2\hbar^2}\Bigg[\\
\nonumber
&&\sum_{k,k',j,j'}g_{kk'}^{(j,j')}\left(2\hat{A}_S^{(k,j)}\hat{\rho}_S(i)\hat{A}_{S}^{(k',j')\dag}-\acomm{\hat{A}_S^{(k',j')}\hat{A}_S^{(k,j)}}{\hat{\rho}_S(i)}\right)\\
\nonumber
&&+\sum_{k,j}g_k^{(j)}\left(2\hat{H}_S\hat{\rho}_S\hat{A}_S^{(k,j)\dag}-\acomm{\hat{A}_S^{(k,j)\dag}\hat{H}_S}{\hat{\rho}_S(i)}\right)\\
\nonumber
&&+\sum_{k,j}g_k^{(j)}\left(2\hat{A}_S^{(j,k)}\hat{\rho}_S(i)\hat{H}_S-\acomm{\hat{H}_S\hat{A}_S^{(k,j)}}{\hat{\rho}_S(i)}\right)\\
\nonumber
&&+\sum_{k,j}\comm{g_{kH_E}^{(j)}\hat{A}_S^{(k,j)}-g_{kH_E}^{(j)*}\hat{A}_S^{(k,j)\dag}}{\hat{\rho}_S(i)}\\
\label{eq:schroedinger_second_order}
&&+\left(2\hat{H}_S\hat{\rho}_S(i)\hat{H}_S-\acomm{\hat{H}_S^2}{\hat{\rho}_S(i)}\right)\Bigg],
%&&\mathcal{L}_2(\hat{\rho}_S(i))=\frac{1}{2\hbar^2}\sum_{k,k',j,j'}g_{kk'}^{(j,j')}\Bigg(2\hat{A}_S^{(k,j)}\hat{\rho}_S(i)\hat{A}_S^{(k',j')\dag}\\
%&&\hspace{70pt}-\acomm{\hat{A}_S^{(k',j')\dag}\hat{A}_S^{(k,j)}}{\hat{\rho}_S(i)}\Bigg),
\ea
where the factors $g_k^{(j)}$, $g_k^{(j,j')}$ and $g_{kH_E}^{(j)}$ are:
\ba
g_k^{(j)}&=&\Tr\left[\hat{B}_{E_ik}^{(j)}\hat{\eta}_{E_i}\right],\\
g_{kk'}^{(j,j')}&=&\Tr\left[\hat{B}_{E_ik'}^{(j')\dag}\hat{B}_{E_ik}^{(j)}\hat{\eta}_{E_i}\right],\\
g_{kH_E}^{(j)}&=&\Tr\left[\hat{H}_{E_{i+1}}\hat{B}_{E_{i+1}k}^{(j)}\hat{\eta}_{E_{i+1}}\right].
\ea
As expected, the term $\mathcal{L}_0$ is simply the identity channel, mapping the density matrix $\hat{\rho}_S$ onto itself.
The term $\mathcal{L}_1$ gives instead a unitary contribution, adding a driving term to the free Hamiltonian $\hat{H}_S$. Finally, the term $\mathcal{L}_2$ describes dissipation phenomena on $S$.
%\begin{widetext}
%In the end, we can write the master equation for system $S$ as:
%\ba
%\nonumber
%\hat{\rho}_S(i+1)=\hat{\rho}_S(i)-\frac{i}{\hbar}g\delta t\sum_{k,j}g_{k}^{(j)}\comm{\hat{A}_S^{(k,j)}}{\hat{\rho}_S(i)}+
%\frac{(g\delta t)^2}{2\hbar^2}\sum_{k,k',j,j'}g_{kk'}^{(j,j')}\Bigg(2\hat{A}_S^{(k,j)}\hat{\rho}_S(i)\hat{A}_S^{(k',j')\dag}-\acomm{\hat{A}_S^{(k',j')\dag}\hat{A}_S^{(k,j)}}{\hat{\rho}_S(i)}\Bigg)\\
%&&\hat{\rho}_S(i+1)=\hat{\rho}_S(i)-\frac{i}{\hbar}g\delta t\comm{\hat{H}_S+\sum_{k,j}g_k^{(j)}\hat{A}_S^{(k,j)}}{\hat{\rho}_S(i)}+\frac{(g\delta t)^2}{2\hbar^2}\Bigg[\sum_{k,k',j,j'}g_{kk'}^{(j,j')}\left(2\hat{A}_S^{(k,j)}\hat{\rho}_S(i)\hat{A}_{S}^{(k',j')\dag}-\acomm{\hat{A}_S^{(k',j')}\hat{A}_S^{(k,j)}}{\hat{\rho}_S(i)}\right)\\
%\nonumber
%&&+\sum_{k,j}g_k^{(j)}\left(2\hat{H}_S\hat{\rho}_S\hat{A}_S^{(k,j)\dag}-\acomm{\hat{A}_S^{(k,j)\dag}\hat{H}_S}{\hat{\rho}_S(i)}\right)+\sum_{k,j}g_k^{(j)}\left(2\hat{A}_S^{(j,k)}\hat{\rho}_S(i)\hat{H}_S-\acomm{\hat{H}_S\hat{A}_S^{(k,j)}}{\hat{\rho}_S(i)}\right)\\
%&&+\sum_{k,j}\comm{g_{kH_E}^{(j)}\hat{A}_S^{(k,j)}-g_{kH_E}^{(j)*}\hat{A}_S^{(k,j)\dag}}{\hat{\rho}_S(i)}+\left(2\hat{H}_S\hat{\rho}_S(i)\hat{H}_S-\acomm{\hat{H}_S^2}{\hat{\rho}_S(i)}\right)\Bigg].
%\ea
%\end{widetext}
The effect of a structured environment can be observed by the presence of the double index $kk'$ in the correlation functions $g_{kk'}^{(j,j')}$, which signal the presence of terms stemming from the common interaction of $S$ with two different subsystems $k,k'$ of the ancilla $E_i$.

Notice also that we did not make the usual assumption $g_k^{(j)}=0$. This is usually a consequence of assuming the environment to be in a thermal state, for which first order correlation functions are identically null. However, since we are now considering a structured environment, even if it is in a thermal state the terms $g_k^{(j)}$ could still be different from zero, due to the coherence between different components of the ancillae.

Let us comment on the continuous limit $\delta t\rightarrow0$. For systems where the first order correlation functions of the environment are null, e.g. ancillae in a thermal state, one can simply 
introduce a dependence of $(\sqrt{\delta t})^{-1}$ in the system-environment interaction Hamiltonian, giving rise to a finite limit~\cite{coll_mod_rev,review_cusumano}. However, using the same trick when the first order correlation functions are non-null, as it could happen in our model, would lead to a divergent equation. This is sometimes solved by inserting a dependence on $\sqrt{\delta t}$ in the coherences of the ancilla state, healing the divergence~\cite{rodrigues,dechiara,hammam}. However, in our model this route is not viable, as there is no way of inserting such an ``ad hoc" dependence. Thus in this work we decided to consider the master equation in Eq.~\eqref{eq:general_me} as a discrete equation without performing the continuous time limit.

This being said, let us now see two examples of a system interacting with a structured bath, in order to better understand when the presence of structured ancillae has effects on the dynamics, and which are those effects.

\begin{figure}[!t]
\centering
\includegraphics[width=\columnwidth]{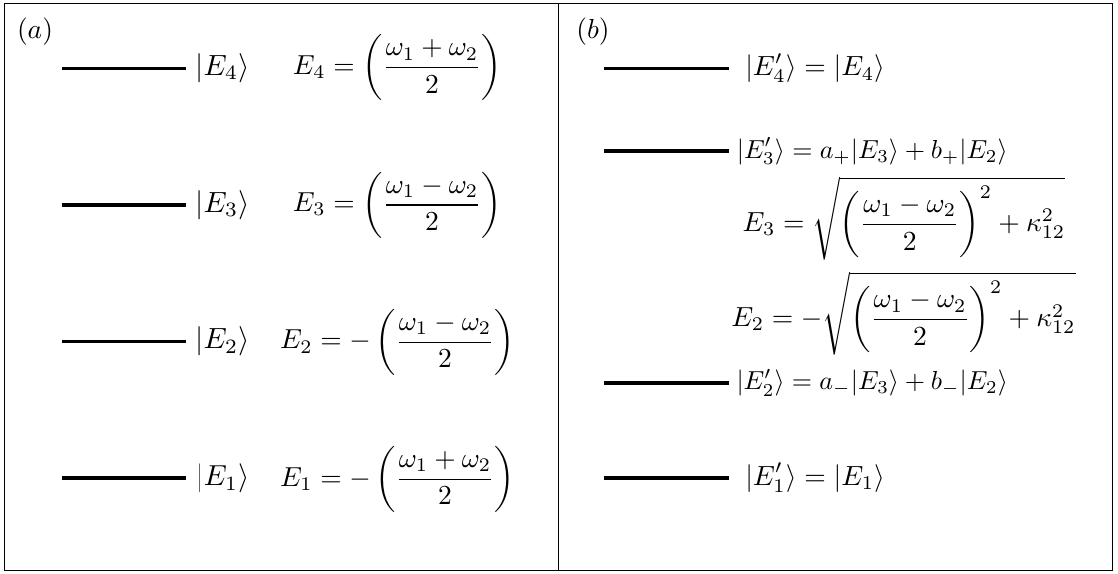}
\caption{Panels (a): Energy levels of two non interacting TLS. Panel (b): Energy levels of two TLS interacting through the Hamiltonian in Eq.~\eqref{eq:interacting_qubits_hamiltonian}. It is immediate to see that the energy levels are changed by the interaction between the two TLS, and that the corresponding eigenstates are linear combinations of the non-interacting eigenstates according to the coefficients $a_{\pm},b_{\pm}$, reported in Eq.~\eqref{eq:a_g_pm}.}
\label{fig:2_qubits_energy_levels}
\end{figure}

%%%%%%%%%%%%%%%%%%%%%%%%%%%%%%%%%%%%%%%%%%%%%
\section{Examples of master equation with structured ancillae\label{sec:examples}}
%%%%%%%%%%%%%%%%%%%%%%%%%%%%%%%%%%%%%%%%%%%%%
\subsection{Example 1\label{sec:example_1}}
As first example we want to consider the case where each ancilla is made out of two interacting TLS, so that the ancilla Hamiltonian reads:
\ba
\label{eq:interacting_qubits_hamiltonian}
\hat{H}_{E_i}=\frac{\omega_1}{2}\hat{\sigma}_1^z+\frac{\omega_2}{2}\hat{\sigma}_2^z+\kappa_{12}\left(\hat{\sigma}_1^+\hat{\sigma}_2^-+\hat{\sigma}_1^-\hat{\sigma}_2^+\right),
\ea
where from now on we set $\hbar=1$ for ease of notation.
Here $\hat{\sigma}_k^z$ is the usual diagonal Pauli matrix, and $\hat{\sigma}_k^\pm=\frac{1}{2}(\hat{\sigma}_k^x\pm i\hat{\sigma}_k^y)$. The bare energy levels of the two TLS are shown in the left panel of Fig.~\ref{fig:2_qubits_energy_levels}. To find the eigenenergies of $\hat{H}_{E_i}$ shown in the right panel of Fig.~\ref{fig:2_qubits_energy_levels} one needs to diagonalize the Hamiltonian in Eq.~\eqref{eq:interacting_qubits_hamiltonian}. To find the eigenenergies one has to solve the equation:
\ba
&&\lambda^4-\lambda^2(x^2+y^2+\kappa_{12}^2)+x^2(y^2+\kappa_{12}^2)=0\\
&&x=\frac{\omega_1+\omega_2}{2},\quad y=\frac{\omega_1-\omega_2}{2}
\ea
whose solutions are:
\ba
&&E'_1=-x=-\frac{\omega_1+\omega_2}{2},\\
&&E'_2=-\sqrt{y^2+\kappa_{12}^2}=-\frac{1}{2}\sqrt{(\omega_1-\omega_2)^2+4\kappa_{12}^2},\quad\\
&&E'_3=\sqrt{y^2+\kappa_{12}^2}=\frac{1}{2}\sqrt{(\omega_1-\omega_2)^2+4\kappa_{12}^2},\\
&&E'_4=+x=\frac{\omega_1+\omega_2}{2}.
\ea
The corresponding eigenvectors are:
\ba
&&\ket{E'_1}=\ket{E_1},\\
&&\ket{E'_2}=\frac{E_2-y}{\sqrt{(E_2-y)^2+\kappa_{12}^2}}\left(\frac{\kappa_{12}}{E_2-y}\ket{E_2}+\ket{E_3}\right)\;\;\\
&&\ket{E'_3}=\frac{E_3-y}{\sqrt{(E_3-y)^2+\kappa_{12}^2}}\left(\frac{\kappa_{12}}{E_3-y}\ket{E_2}+\ket{E_3}\right)\\
&&\ket{E'_4}=\ket{E_4}.
\ea
Note that the weights in $\ket{E'_2},\ket{E'_3}$ depend on the detuning $y$. If the two TLS have the same frequency, i.e. $\omega_1=\omega_2$, then $\ket{E'_2}$ and $\ket{E'_3}$ become a singlet and a triplet state with zero magnetization respectively.

We now assume that the open system $S$ is also a TLS with free Hamiltonian:
\ba
\label{eq:system_hamiltonian_ex_2}
\hat{H}_S=\frac{\omega_S}{2}\hat{\sigma}_S^z,
\ea
interacting with the ancillae through the interaction Hamiltonian:
\ba
\hat{H}_{SE_i}=\alpha_1(\hat{\sigma}_S^+\hat{\sigma}_1^-+\hat{\sigma}_S^-\hat{\sigma}_1^+)+\alpha_2(\hat{\sigma}_S^+\hat{\sigma}_2^-+\hat{\sigma}_S^-\hat{\sigma}_2^+),
\ea
where in the right hand side (rhs) of the equation we have dropped for simplicity the index $i$ and the operators $\hat{\sigma}_{1,2}^\pm$ are the same appearing in Eq.~\eqref{eq:interacting_qubits_hamiltonian}.
Moreover, we assume the ancillae to be initially in a thermal state at inverse temperature $\beta$, that is:
\ba
\hat{\eta}_{E_i}=\frac{e^{-\beta\hat{H}_{E_i}}}{Z},\;Z=\Tr[e^{-\beta\hat{H}_{E_i}}].
\ea

We first look at the possible unitary contributions, i.e. the ones stemming from the terms proportional to $g_k^{(j)}$ or $g_{kH_E}^{(j)}$. As shown in App.~\ref{app:example_1} all these terms are zero, and thus there is no unitary contribution to the dynamics in this case.

Next we look at the structure of the dissipative part of the dynamics, the one described by the term $\mathcal{L}_2(\hat{\rho}_S)$.
To do this, we give a look at the correlation functions $g_{kk'}^{j,j'}$ of the environment, starting with the ones where $k=k'$. There are four non-zero such correlation functions:
\ba
g_{11}^{(+,-)}=\Tr\left[\hat{\sigma}_1^-\hat{\sigma}_1^+\hat{\eta}_{E_i}\right],
g_{11}^{(-,+)}=\Tr\left[\hat{\sigma}_1^+\hat{\sigma}_1^-\hat{\eta}_{E_i}\right],\\
g_{22}^{(+,-)}=\Tr\left[\hat{\sigma}_2^-\hat{\sigma}_2^+\hat{\eta}_{E_i}\right],
g_{22}^{(-,+)}=\Tr\left[\hat{\sigma}_2^+\hat{\sigma}_2^-\hat{\eta}_{E_i}\right],
\ea
which are computed explicitly in App.~\ref{app:example_1}.
%This correlation functions give rise to the usual thermal dissipators. For instance $g_{11}^{(+,-)}$ is the factor of the term:
%\ba
%\label{eq:01}
%g_{11}^{(+,-)}\left(2\hat{\sigma}_S^-\hat{\rho}_S\hat{\sigma}_S^+-\acomm{\hat{\sigma}_S^+\hat{\sigma}_S^-}{\hat{\rho}_S}\right).
%\ea
In addition to this, we also have cross correlation functions, i.e. those with $k\neq k'$. In this case we have eight possible terms, which are complex conjugates in couples:
\ba
\label{eq:g_12pm}
&&g_{12}^{(+,-)}=g_{21}^{(+,-)*}=\Tr\left[\hat{\sigma}_1^-\hat{\sigma}_2^+\hat{\eta}_{E_i}\right]\\
\label{eq:g_12mp}
&&g_{12}^{(-,+)}=g_{21}^{(-,+)*}=\Tr\left[\hat{\sigma}_1^+\hat{\sigma}_2^-\hat{\eta}_{E_i}\right]\\
&&g_{12}^{(-,-)}=g_{21}^{(-,-)*}=\Tr\left[\hat{\sigma}_1^+\hat{\sigma}_2^+\hat{\eta}_{E_i}\right]\\
&&g_{12}^{(+,+)}=g_{21}^{(+,+)*}=\Tr\left[\hat{\sigma}_1^-\hat{\sigma}_2^-\hat{\eta}_{E_i}\right].
\ea

As shown in App.~\ref{app:example_1}, the only non zero elements are the ones in Eqs.(~\ref{eq:g_12pm},~\ref{eq:g_12mp}). Conversely, the terms with $g_{12}^{(+,+)}$ and $g_{12}^{(-,-)}$ are equal to zero. In the end, we can write the master equation for the dynamics of $S$ in the interaction picture as follows:
\ba
%\nonumber
%&&\hat{\rho}_S(i+1)=\hat{\rho}_{S}(i)\\
%\nonumber
%&&+\frac{(g\delta t)^2}{2}\Bigg\{\left(g_{11}^{(+,-)}+g_{22}^{(+,-)}+g_{12}^{(+,-)}+g_{21}^{(+,-)}\right)\\
%\nonumber
%&&\left[2\hat{\sigma}_S^-\hat{\rho}_S(i)\hat{\sigma}_S^+-\acomm{\hat{\sigma}_S^+\hat{\sigma}_S^-}{\hat{\rho}_S(i)}\right]\\
%\nonumber
%&&+\left(g_{11}^{(-,+)}+g_{22}^{(-,+)}+g_{12}^{(-,+)}+g_{21}^{(-,+)}\right)\\
%&&\left[2\hat{\sigma}_S^+\hat{\rho}_S(i)\hat{\sigma}_S^--\acomm{\hat{\sigma}_S^-\hat{\sigma}_S^+}{\hat{\rho}_S(i)}\right]\Bigg\}
\nonumber
&&\hat{\rho}_S(i+1)=\hat{\rho}_S(i)-i\delta t\comm{\hat{H}_S}{\hat{\rho}_S(i)}\\
\nonumber
&&+\frac{\delta t^2}{2}\Bigg[\sum_{i,j=1,2}g_{ij}^{(+,-)}\mathcal{D}[\hat{\sigma}_S^+](\hat{\rho}_S(i))\\
%\nonumber
%&&\left(2\hat{\sigma}_S^+\hat{\rho}_S(i)\hat{\sigma}_S^--\acomm{\hat{\sigma}_S^-\hat{\sigma}_S^+}{\hat{\rho}_S(i)}\right)\\
\label{eq:2_qubits_ancillas_me}
&&\sum_{i,j=1,2}g_{ij}^{(-,+)}\mathcal{D}[\hat{\sigma}_S^-](\hat{\rho}_S(i))+\frac{\omega_S^2}{4}\mathcal{D}[\hat{\sigma}_S^z](\hat{\rho}_S(i))\Bigg],%+\left(g_{11}^{(-,+)}+g_{22}^{(-,+)}+g_{12}^{(-,+)}+g_{21}^{(-,+)}\right)\\
%\nonumber
%&&(2\hat{\sigma}_S^-\hat{\rho}_S(i)\hat{\sigma}_S^+-\acomm{\hat{\sigma}_S^+\hat{\sigma}_S^-}{\hat{\rho}_S(i)})\\
%\label{eq:2_qubits_ancillas_me}
%&&+\frac{\omega_S^2}{4}\left(2\hat{\sigma}_S^z\hat{\rho}_S(i)\hat{\sigma}_S^z-2\hat{\rho}_S(i)\right)\Bigg].
\ea
where we have defined:
\ba
\mathcal{D}[\hat{A}](\cdots)=2\hat{A}\cdots\hat{A}^\dag-\acomm{\hat{A}^\dag\hat{A}}{\cdots},\\
\mathcal{D}'[\hat{A},\hat{B}](\cdots)=2\hat{A}\cdots\hat{B}-\acomm{\hat{B}\hat{A}}{\cdots}.
\ea
This is the usual discrete Markovian master equation for a system interacting with a thermal bath. The presence of coherence in the thermal bath has no effect on the dynamics and thus the system will simply reach thermal equilibrium at the steady state, without the formation of any coherence. One can in fact compute explicitly the steady state satisfying $\hat{\rho}_S(i+1)=\hat{\rho}_S(i)$ to obtain:
\ba
\hat{\rho}_{S}^{steady}=\frac{1}{\sum_{i,j,\pm}g_{ij}^{(\pm,\mp)}}
\begin{bmatrix}
\sum_{i,j}g_{ij}^{(-,+)}&0\\
0&\sum_{i,j}g_{ij}^{(+,-)}
\end{bmatrix}.
\ea

With this example we can see that when system $S$ interacts with each qubit of the ancilla independently, then the presence of coherence in the bath degrees of freedom has no effects on the dynamics. In the next example we are going to see how, when the interaction between the system and the ancilla involves joint degrees of freedom of the ancilla, i.e. degrees of freedom that have coherence, then the presence of coherence becomes significant for the dynamics.

%%%%%%%%%%%%%%%%%%%%%%%%%%%%%%%%%%%%%%%%%%%%%%%%%%%%%%%
\subsection{Example 2\label{sec:example_2}}
%%%%%%%%%%%%%%%%%%%%%%%%%%%%%%%%%%%%%%%%%%%%%%%%%%%%%%%
As a second example we consider the same system we examined in Sec.~\ref{sec:example_1}, but this time with an interaction Hamiltonian between the system and the ancillae of the form:
\ba
\label{eq:int_hamiltonian_ex_2}
\hat{H}_{SE_i}=\alpha\hat{\sigma}_S^+\hat{\sigma}_{E_i}^{-}+\alpha^*\hat{\sigma}_S^-\hat{\sigma}_{E_i}^+
\ea
where the operators $\hat{\sigma}_{E_i}^\pm$ are defined as:
\ba
\hat{\sigma}_{E_i}^+=\dyad{E_3}{E_2}\quad\hat{\sigma}_{E_i}^-=\dyad{E_2}{E_3},
\ea
and the states $\ket{E_2},\ket{E_3}$ are the bare levels of the ancilla Hamiltonian.
%\ba
%\ket{E_2}=\ket{\downarrow\uparrow},\;\ket{E_3}=\ket{\uparrow\downarrow}.
%\ea
Thanks to the results of the previous section we can immediately write the Hamiltonian and dissipative parts of the dynamics. We have:
\ba
\nonumber
\mathcal{L}_1&=&-i\comm{\left(g_{E_i}^{(+)}\right)\hat{\sigma}_S^-+\left(g_{E_i}^{(-)}\right)\hat{\sigma}_S^+}{\hat{\rho}_S}\\
&=&-i\comm{\Re{g_{E_i}^{(+)}}\hat{\sigma}_S^x+\Im{g_{E_i}^{(+)}}\hat{\sigma}_S^y}{\hat{\rho}_S},
\ea
where the constants $g_{E_i}^{(\pm)}$ are:
\ba
g_{E_i}^{(+)}=\Tr\left[\hat{\sigma}_{E_i}^+\hat{\eta}_{E_i}\right]=g_{E_i}^{(-)*}.
\ea
Note that for an unstructured environment these terms would be identically zero, as they are proportional to the amount of coherence between the levels $\ket{E_2}$ and $\ket{E_3}$.
In practice, the term $\mathcal{L}_1(\hat{\rho}_S)$ describe the unitary evolution due to an effective Hamiltonian of the form:
\ba
\nonumber
\hat{H}_{S}^{\rm eff}=\Re{g_{E_i}^{(+)}}\hat{\sigma}_S^x+\Im{g_{E_i}^{(+)}}\hat{\sigma}_S^y.
\ea

\begin{figure*}[!t]
\centering
\includegraphics[width=\textwidth]{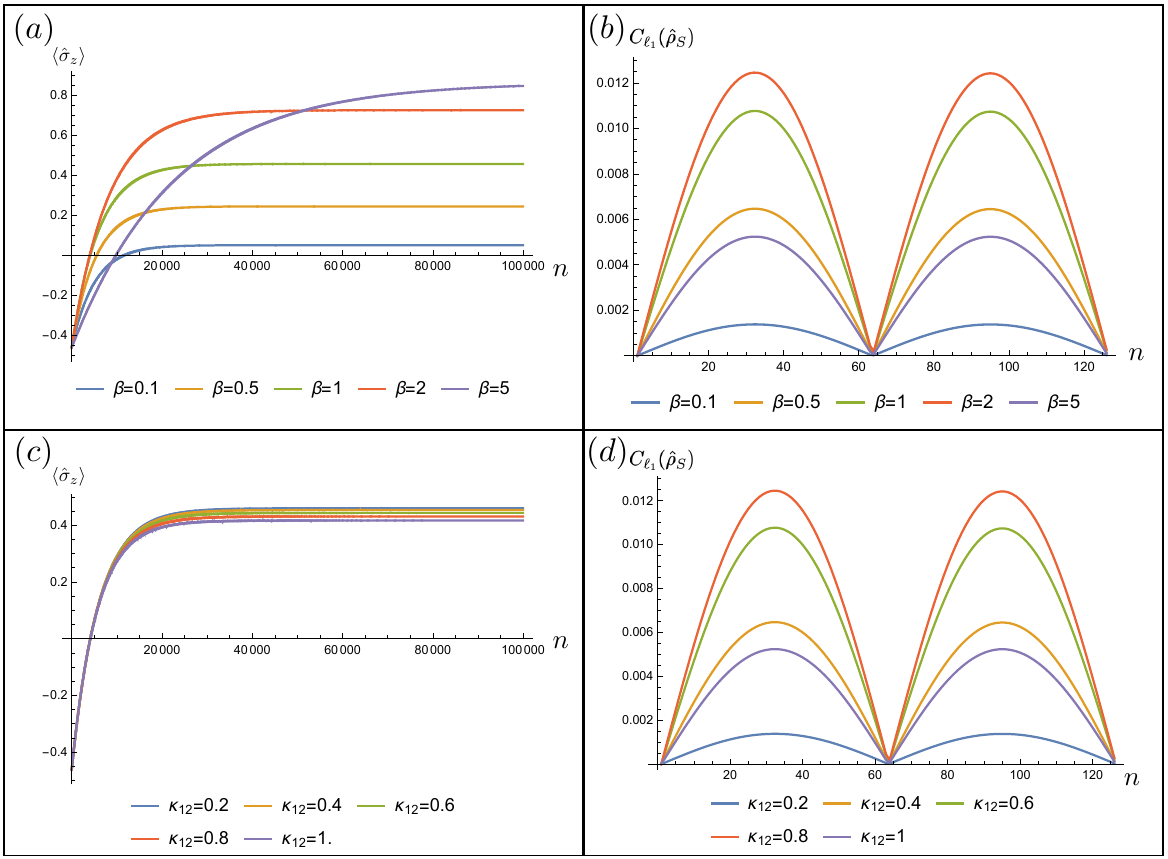}
\caption{Plots of the evolution of populations and coherence. In all the plots $\delta t=0.1$, $\omega_S=1$, $\omega_1=0.5$, $\omega_2=1.5$, $\kappa_{12}=0.3$, $\alpha=0.1$. The initial state is a thermal state of the form $\hat{\rho}_S(0)=\exp[-\beta_S\hat{H}_S]/Z_S$, $Z_S$ being the partition function and $\beta_S=1$. Panel (a): plot of the expectation value $\langle\hat{\sigma}_z\rangle$ as a function of the number of collisions. As one can see, the population tends to equilibrate towards the steady state value, which is not simply the thermal population, as the steady state also has coherence. This observation in particular justifies the behavior of the population for the case $\beta=\beta_S$, which without the creation of coherence should just be a line. In panels (b) the $\ell_1$ norm of coherence $C_{\ell_1}(\hat{\rho}_S)$ is plotted. One can observe how the amount of coherence is maximal when the system and the environment have the same temperature. In panel (c) and (d) the same quantities are plotted, this time for different values of $\kappa_{12}$, the environmental temperature being $\beta=\beta_S=1$. As one can observe, there is a small difference in the equilibrium populations, due to the fact that the amount of coherence created is different. Moreover, one can observe that the stronger the interaction between the two TLSs composing the ancilla, the greater the amount of coherence created.}
\label{fig:example_2_dynamics}
\end{figure*}

While we leave the explicit calculations of the coefficients to App.~\ref{sec:appendix_2}, we can immediately write down the master equation as:
\ba
\nonumber
&&\hat{\rho}_S(i+1)=\hat{\rho}_S(i)-i\delta t\comm{\hat{H}_S+g_{E_i}^{(-)}\hat{\sigma}_S^++g_{E_i}^{(+)}\hat{\sigma}_S^-}{\hat{\rho}_S(i)}\\
\nonumber
&&+\frac{\delta t^2}{2}\Bigg[\frac{\omega_S^2}{4}\mathcal{D}[\hat{\sigma}_S^z](\hat{\rho}_S(i))+g_{E_i}^{(+,-)}\mathcal{D}[\hat{\sigma}_S^-](\hat{\rho}_S(i))\\
\nonumber
&&+g_{E_i}^{(-,+)}\mathcal{D}[\hat{\sigma}_S^-](\hat{\rho}_S(i))+\frac{g_{E_i}^{(+)}\omega_S}{2}\mathcal{D}'[\hat{\sigma}_S^-,\hat{\sigma}_S^z](\hat{\rho}_S(i))\\
\nonumber
&&+\frac{g_{E_i}^{(+)}\omega_S}{2}\mathcal{D}'[\hat{\sigma}_S^z,\hat{\sigma}_S^-](\hat{\rho}_S(i))+\frac{g_{E_i}^{(-)}\omega_S}{2}\mathcal{D}'[\hat{\sigma}_S^+,\hat{\sigma}_S^z](\hat{\rho}_S(i))\\
\label{eq:master_equation_2}
&&\qquad+\frac{g_{E_i}^{(-)}\omega_S}{2}\mathcal{D}'[\hat{\sigma}_S^z,\hat{\sigma}_S^+](\hat{\rho}_S(i))\Bigg].
\ea
The dynamics described by this master equation consists in the usual dissipative contribution and dephasing term driving the system towards the thermal equilibrium, plus a unitary contribution which creates coherence in the system.
On one hand one can see that the populations, i.e. $\rho_{11}$ and $\rho_{22}$, evolve according to the usual term due to dissipation, plus another term due to the presence of the unitary driving. On the other hand, the coherences $\rho_{12}$ and $\rho_{21}$, besides the usual decay term due to the presence of the environment, evolves also according to terms due to the presence of the unitary driving, thus leading to a non-zero value in the steady state.

In fact it is possible to compute the steady state of system $S$ by imposing $\hat{\rho}(i+1)=\hat{\rho}(i)$, obtaining analytical expressions for $\langle\hat{\sigma}_x\rangle$, $\langle\hat{\sigma}_x\rangle$, $\langle\hat{\sigma}_z\rangle$, which are reported in App.~\ref{sec:appendix_2}.
What can be seen from these expressions is that the coherence is non null in the steady state, as both $\langle\hat{\sigma}_x\rangle$ and $\langle\hat{\sigma}_y\rangle$ are non-zero. Moreover, it is worth noticing that the equilibrium populations will not simply be the thermal populations, due to the fact that coherence is created in the system $S$.

\begin{figure}[!t]
\centering
\includegraphics[width=\columnwidth]{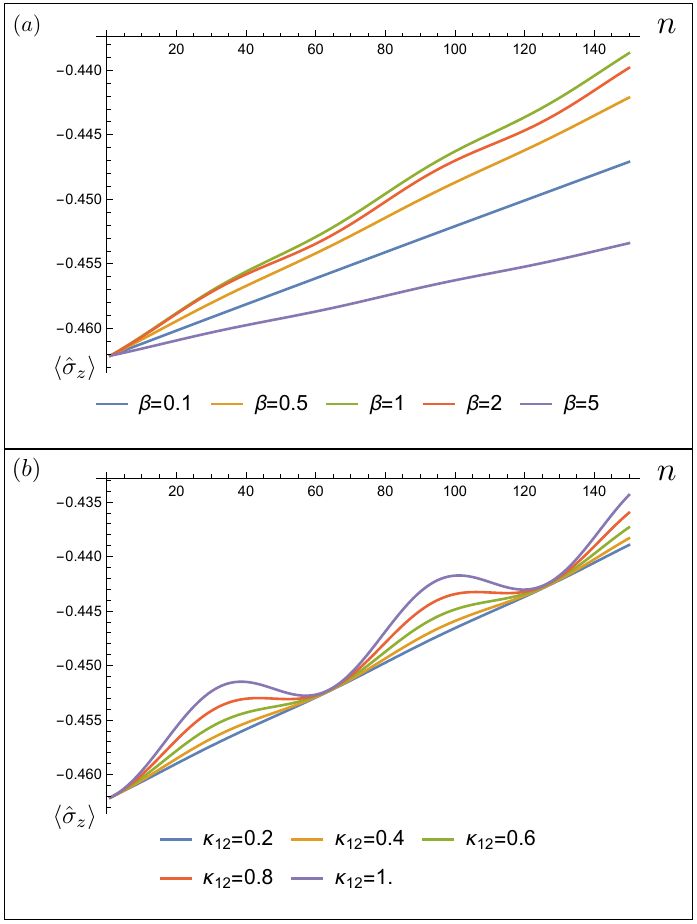}
\caption{Detail of the evolution of $\langle\hat{\sigma}_z\rangle$. Panel (a): plot of the evolution of $\langle\hat{\sigma}_z\rangle$ for different values of the environmental temperature $\beta$. The values of the other parameters are $\delta t=0.1$, $\omega_S=1$, $\omega_1=0.5$, $\omega_2=1.5$, $\kappa_{12}=0.3$, $\alpha=0.2$. It can be observed that, while going towards its steady state value, $\langle\hat{\sigma}_z\rangle$ oscillates due to the presence of the unitary term. Panel (b): evolution of $\langle\hat{\sigma}_z\rangle$ for different values of $\kappa_{12}$, where the environmental temperature is $\beta=\beta_S=1$. From this plot one can see that the amplitude of the oscillations of $\langle\hat{\sigma}_z\rangle$ is larger for larger values of $\kappa_{12}$.}
\label{fig:populations_oscillations}
\end{figure}

We now want to analyze with more detail the dynamics of $S$, in particular checking how the populations and the coherence evolve, that is, we want to look at the transient dynamics before reaching the steady state. In order to quantify the amount of coherence in the state $\hat{\rho}_S$ we use the $\ell_1$ norm of coherence~\cite{plenio} $C_{\ell_1}(\hat{\rho})$, defined as:
\ba
C_{\ell_1}(\hat{\rho})=\sum_{i\neq j}\left|\hat{\rho}_{ij}\right|,
\ea
which in the present example is simply evaluated as:
\ba
C_{\ell_1}(\hat{\rho}_S)=\left|\langle\hat{\sigma}_x\rangle+i\langle\hat{\sigma}_y\rangle\right|.
\ea

In Fig.~\ref{fig:example_2_dynamics} some plots for the populations $\langle\hat{\sigma_z}\rangle$ and $C_{\ell_1}(\hat{\rho}_S)$ are shown. In the upper plots the evolution of these quantities is shown for different values of the environmental temperature $\beta$, while in the bottom plots the same quantities are plotted for different values of $\kappa_{12}$.

Some observations are to be made. First, one can notice that the equilibrium populations are not the thermal ones, consistently with the fact that coherence is created in system $S$. Furthermore, one can observe oscillations in the value of $C_{\ell_1}(\hat{\rho}_S)$. These oscillations are slowly suppressed as the system reaches its steady state. Finally, one can clearly observe that the total amount of coherence available to system $S$ is strongly dependent on the amount of coherence in the ancillae, ultimately depending on the value of $\kappa_{12}$.

\begin{figure}[!t]
\centering
\includegraphics[width=\columnwidth]{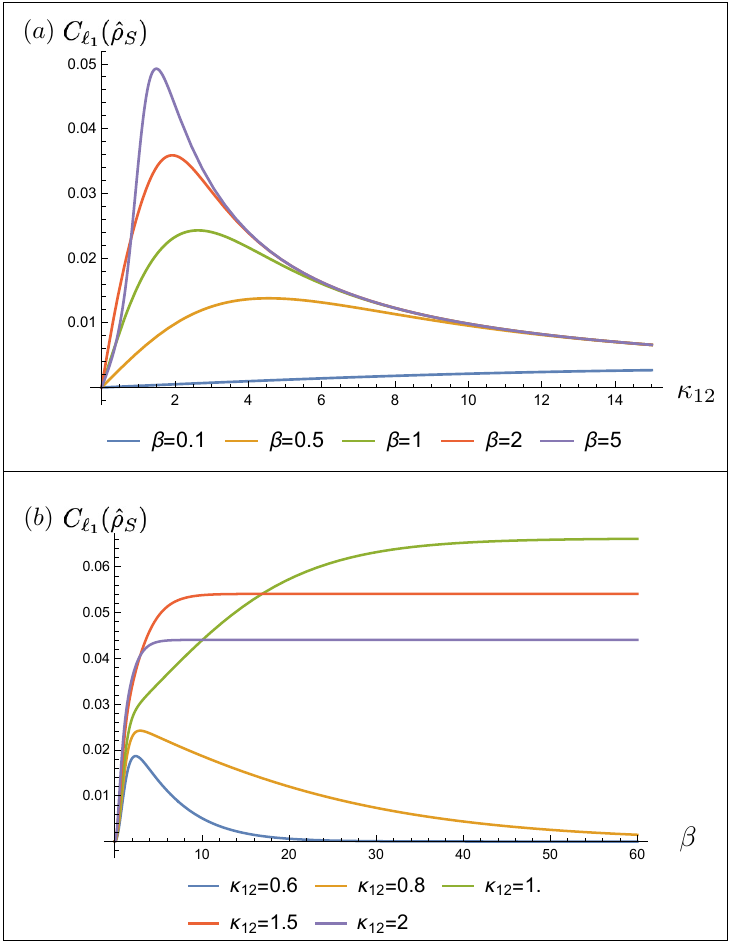}
\caption{Plots of $C_{\ell_1}(\hat{\rho}_S)$ at the steady state. The values of the parameters are $\delta t=0.1$, $\omega_S=1$, $\omega_1=0.5$, $\omega_2=1.5$, $\alpha=0.2$ Panel (a):  plot of $C_{\ell_1}(\hat{\rho}_S)$ as a function of $\kappa_{12}$ for different values of $\beta$. In this case it can be noticed that in general the amount of coherence generated in the steady state is larger for lower temperatures. However the behavior of the coherence for large values of $\kappa_{12}$ shows that the amount of coherence reaches a steady value independently of the temperature. Moreover, one can see that for specific values of $\kappa_{12}$ the amount of coherence is larger.  Panel (b): plot of $C_{\ell_1}(\hat{\rho}_S)$ as a function of $\beta$ for different values of $\kappa_{12}$. On one hand, it can be immediately noticed that, for values of $\kappa_{12}\lesssim1$ one has larger values of $C_{\ell_1}(\hat{\rho}_S)$ for larger values of $\kappa_{12}$. On the other hand, for values of $\kappa_{12}\gtrsim1$, in the limit of very large $\beta$, the coherence reaches always the same steady value. The case $\kappa_{12}\simeq1$ represents the border between the two scenarios, where the coherence, in the limit of large $\beta$ reaches the higher value.}
\label{fig:sigma_x_steady}
\end{figure}

Moreover, in Fig.~\ref{fig:populations_oscillations} we have plotted a detail of the evolution of the populations for various temperatures $\beta$ and different $\kappa_{12}$. It can be observed, consistently with the presence of the unitary driving term, that while going to their steady state value, the populations also oscillates. One can also observe that these oscillations become wider for larger values of $\kappa_{12}$, as the strength of the unitary driving becomes larger.

In Fig.~\ref{fig:sigma_x_steady} we have plotted the steady state value of $C_{\ell_1}(\hat{\rho}_S)$ as a function of $\kappa_{12}$ for several values of the environmental temperature $\beta$ and as a function of $\beta$ for different values of $\kappa_{12}$. As one can see in Fig.~\ref{fig:sigma_x_steady}(a), on one hand for all values of $\beta$, as $\kappa_{12}$ grows large the steady state coherence converges towards an asymptotic value. On the other hand, one can observe a peak in $C_{\ell_1}(\hat{\rho}_S)$ for specific values of $\kappa_{12}$ and $\beta$. In Fig.~\ref{fig:sigma_x_steady}(b) one can instead observe a different behavior as a function of the temperature for different values of $\kappa_{12}$. In fact, for $\kappa_{12}\lesssim1$ it can be observed that the steady state coherence goes to zero as $\beta$ grows large. Conversely, for values of $\kappa_{12}\gtrsim1$, as $\beta$ grows large one can observe that the steady state coherence goes towards a non zero value, the largest value being obtained for $\kappa_{12}\simeq1$. In conclusion, the dependence of $C_{\ell_1}(\hat{\rho}_S)$ on $\kappa_{12}$ and $\beta$ is in general non trivial, though its main characteristics can be extrapolated. These plots demonstrate the result we anticipated and one of the main contributions of this paper: the generation of steady-state coherence using only thermal resources.

%While the presence of coherence in the steady state is already significant per se, in Sec.~\ref{sec:thermodynamics} we are going to see how the creation of coherence in system $S$ influences the thermodynamics of system $S$.

%%%%%%%%%%%%%%%%%%%%%%%%%%%%%%%%%%%%%%%%
\section{Thermodynamics\label{sec:thermodynamics}}
%%%%%%%%%%%%%%%%%%%%%%%%%%%%%%%%%%%%%%%%
In this section we want to analyze the thermodynamics of a system when interacting with a bath made out of structured ancillae. We will first define the relevant thermodynamic quantities, such as internal energy, work and heat. Then we will make some general considerations based on the general form of the master equation before moving to a practical example based on the system described in Sec.~\ref{sec:example_2}. We will show that the creation of coherence in the steady state requires work to be injected into the system, thus recovering the first principle of thermodynamics in its standard form and fulfilling the second principle of thermodynamics.

\subsection{Defining the thermodynamic quantities\label{sec:thermo_quantities}}

We define the thermodynamic quantities starting from the internal energy of system $S$, which is simply:
\ba
U=\Tr\left[\hat{H}_S\hat{\rho}_S\right],
\ea
so that at each step the internal energy variation can be defined as:
\ba
\nonumber
\Delta U(i+1)&=&\Tr\left[\hat{H}_S\left(\hat{\rho}_S(i+1)-\hat{\rho}_S(i)\right)\right]\\
\label{eq:int_en_var_def}
&=&\Tr\left[\hat{H}_S\left(\hat{U}_{SE_{i+1}}\hat{R}_{S\cal E}(i)\hat{U}_{SE_{i+1}}-\hat{R}_{S\cal E}(i)\right)\right].\qquad
\ea
By inserting the expressions in Eqs.~(~\ref{eq:schroedinger_fo},~\ref{eq:schroedinger_so}) into Eq.~\eqref{eq:int_en_var_def} we obtain:
\begin{widetext}
\ba
\nonumber
&&\Delta U(i+1)=\Tr\left[\hat{H}_S\mathcal{U}_1(\hat{R}_{S\cal E}(i))\right]+\Tr\left[\hat{H}_S\mathcal{U}_2(\hat{R}_{S\cal E}(i))\right]\\
&&=-i\delta t\Tr\left[\comm{\hat{H}_S}{\hat{H}_{SE_{i+1}}}\hat{R}_{S\cal E}(i)\right]+\frac{\delta t^2}{2}\Tr\left[\comm{\hat{H}_S+\hat{H}_{E_{i+1}}+\hat{H}_{SE_{i+1}}}{\comm{\hat{H}_S}{\hat{H}_{SE_{i+1}}}}\hat{R}_{S\cal E}(i)\right].\label{eq:int_en_var_explicit}
\ea
\end{widetext}

\begin{figure*}[!t]
\centering
\includegraphics[width=\textwidth]{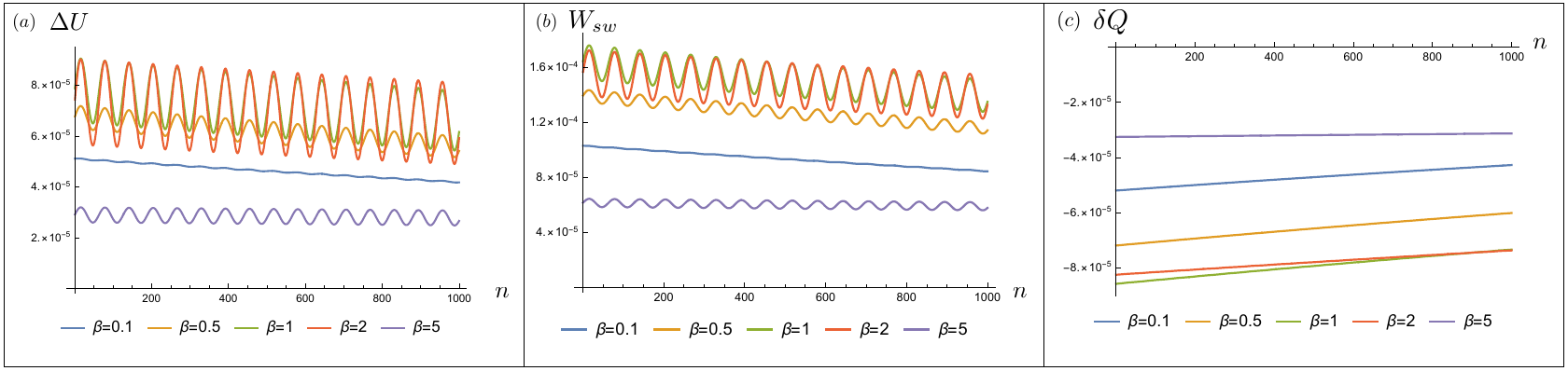}
\caption{Plot of $\Delta U$, $\delta Q$, $W_{sw}$ for the system in Example 2 and the same parameters value s in Fig.~\ref{fig:example_2_dynamics}. Panel (a), (b), (c): plots of the three quantities for the initial part of the dynamics. As one can see, as both the internal energy change and the switching work depend on the coherences in system $S$, one can observe oscillations in these quantities. Conversely, heat does not oscillate, as it only depends on the populations of $S$ and the ancilla matrix elements.}
% Panel (d), (e), (f): plots of the three quantities for the final part of the dynamics. As one can see, for the parameters regime for which the systems arrives to its steady state, all the quantities stop oscillating, the variation of internal energy $\Delta U$ being zero and the switching work $W_{sw}$ being thus equal to $-\delta Q$.}
\label{fig:thermo_plots}
\end{figure*}

We now define the heat exchanged by the system, assumed to be positive when it is absorbed by the system.
At each interaction between the system and an ancilla, we thus define the heat exchanged as the opposite of the energy variation of the ancilla:
\ba
\nonumber
&&\delta Q(i+1)=-\Tr\left[\hat{H}_{E_i}(\hat{\eta}_{E_{i+1}}^{out}-\hat{\eta}_{E_{i+1}}^{in})\right]\quad\\
\label{eq:heat_definition}
&&=-\Tr\left[\hat{H}_{E_{i+1}}\left(\hat{U}_{SE_{i+1}}\hat{R}_{S\cal E}(i)\hat{U}_{SE_{i+1}}^\dag-\hat{R}_{S\cal E}(i)\right)\right],\qquad
\ea
where we have defined:
\ba
\hat{\eta}_{E_{i+1}}^{out}=\Tr_S\left[\hat{U}_{SE_{i+1}}\hat{R}_{S\cal E}(i)\hat{U}_{SE_{i+1}}^\dag\right].
\ea
Definition~\eqref{eq:heat_definition} is justified since the initial state of the environment is thermal.
Once again we can insert the expressions in Eqs.~(~\ref{eq:schroedinger_fo},~\ref{eq:schroedinger_so}) to obtain:
\begin{widetext}
\ba
\nonumber
&&\delta Q(i+1)=-\Tr\left[\hat{H}_{E_{i+1}}\mathcal{U}_1(\hat{R}_{S\cal E}(i))\right]+\Tr\left[\hat{H}_{E_{i+1}}\mathcal{U}_2(\hat{R}_{S\cal E})\right]\\
\nonumber
&&=i\delta t\Tr\left[\comm{\hat{H}_{E_{i+1}}}{\hat{H}_{SE_{i+1}}}\hat{R}_{S\cal E}(i)\right]-\frac{\delta t^2}{2}\Tr\left[\comm{\hat{H}_S+\hat{H}_{E_{i+1}}+\hat{H}_{SE_{i+1}}}{\comm{\hat{H}_{E_{i+1}}}{\hat{H}_{SE_{i+1}}}}\hat{R}_{S\cal E}(i)\right],\label{eq:heat_explicit}
\ea
\end{widetext}
Moreover, any variation of the system Hamiltonian will contribute to work, so that we define:
\ba
\delta W=\Tr[\Delta\hat{H}_S\hat{\rho}_S].
\ea
As a consequence, we also have to take into consideration the switching work contribution~\cite{strasberg}, namely the one stemming from turning on and off of the interaction between the system and the ancilla, which reads:
\ba
\nonumber
W_{sw}(i+1)&=&\Tr\left[\hat{H}_{SE_i}\left(\hat{R}_{S\cal E}(i)-\hat{U}_{SE_{i+1}}\hat{R}_{S\cal E}(i)\hat{U}_{SE_{i+1}}^\dag\right)\right],\\\label{eq:switching_work}
\ea
\begin{widetext}
Also in this case we can use the expressions in Eqs.~(~\ref{eq:schroedinger_fo},~\ref{eq:schroedinger_so}) and obtain:
\ba
\label{eq:switching_work_explicit}
&&W_{sw}(i+1)=-\Tr\left[\hat{H}_{SE_{i+1}}\mathcal{U}_1(\hat{R}_{S\cal E}(i))\right]-\Tr\left[\hat{H}_{SE_{i+1}}\mathcal{U}_2(\hat{R}_{S\cal E}(i))\right]=i\delta t\Tr\left[\comm{\hat{H}_{SE_{i+1}}}{\hat{H}_S+\hat{H}_{E_{i+1}}}\hat{R}_{S\cal E}(i)\right]\\
\nonumber
&&-\frac{\delta t^2}{2}\Tr\left[\comm{\hat{H}_S+\hat{H}_{E_{i+1}}+\hat{H}_{SE_{i+1}}}{\comm{\hat{H}_{SE_{i+1}}}{\hat{H}_S}}\hat{R}_{S\cal E}(i)\right]-\frac{\delta t^2}{2}\Tr\left[\comm{\hat{H}_S+\hat{H}_{E_{i+1}}+\hat{H}_{SE_{i+1}}}{\comm{\hat{H}_{SE_{i+1}}}{\hat{H}_{E_{i+1}}}}\hat{R}_{S\cal E}(i)\right].
\ea
\end{widetext}
It is important to note that this term will be different from zero in case the interaction Hamiltonian is energy non-conserving. Note also that in our convention the work is positive when it is absorbed by the system and negative when it is extracted.

Putting together Eqs.(~\ref{eq:int_en_var_explicit},~\ref{eq:heat_explicit},~\ref{eq:switching_work_explicit}) we can observe that the first principle of thermodynamics is automatically fulfilled, as it reads:
\ba
\Delta U=\delta Q+\delta W+W_{sw},
\ea
where in absence of external driving $\delta W=0$.
The interpretation of Eqs.(~\ref{eq:int_en_var_explicit},~\ref{eq:heat_explicit},~\ref{eq:switching_work_explicit}) is straightforward as one can actually perform explicitly the traces, obtaining the expressions reported in App.~\ref{sec:appendix_3}.. At each collision, work is needed in order to turn on and off the interaction between the system and the ancilla. Part of this work is then dissipated as heat into the bath, while another part is used to change the internal energy of the system.

Let us now turn our attention to the entropic quantities. The von Neumann entropy of a quantum state is defined as:
\ba
S(\hat{\rho}_S)=-\Tr\left[\hat{\rho}_S\log\hat{\rho}_S\right].
\ea
We thus define the system $S$ entropy variation during one collision as:
\ba
\Delta S_S(i+1)=S(\hat{\rho}_S(i+1))-S(\hat{\rho}_S(i)).
\ea
%while the entropy variation of the environment is defined as:
%\ba
%\Delta S_{E_{i+1}}=S(\hat{\eta}_{E_{i+1}}^{out})-S(\hat{\eta}_{E_{i+1}}^{in}).
%\ea
%Thus, the total entropy increase during one collision is:
%\ba
%\Delta S_{\rm tot}(i+1)=\Delta S_S(i+1)+\Delta S_{E_{i+1}}
%\ea
The entropy production, describing the total amount of entropy created during one collision, including the contributions due to the change of the ancilla state and the one due to the loss of information, can be defined as~\cite{strasberg}:
\ba
\Sigma(i)=\mathcal{I}(\hat{\rho}_S(i);\hat{\eta}_{E_{i}}^{out})+S(\hat{\eta}_{E_{i}}^{out}||\hat{\eta}_{E_{i}}^{in}),
\ea
where
\ba
\nonumber
&&\mathcal{I}(\hat{\rho}_S(i);\hat{\eta}_{E_{i}}^{out})=\\
\nonumber
&&=S(\hat{\rho}_S(i))+S(\hat{\eta}_{E_i}^{out})-S(\hat{U}_{SE_i}R_{S\cal E}(i)\hat{U}_{SE_{i}})\\
\nonumber
&&=S(\hat{\rho}_S(i))+S(\hat{\eta}_{E_i}^{out})-S(\hat{\rho}_S(i-1))-S(\hat{\eta}_{E_i}^{in}),
\ea
is the mutual information between system $S$ and the ancilla $E_i$ after their interaction, while:
\ba
S(\hat{\eta}_{E_{i}}^{out}||\hat{\eta}_{E_{i}}^{in})=\Tr\left[\hat{\eta}_{E_i}^{out}\log\hat{\eta}_{E_i}^{out}\right]-\Tr\left[\hat{\eta}_{E_i}^{out}\log\hat{\eta}_{E_i}^{in}\right],
\ea
is the relative entropy between the state of the ancilla after the interaction and the one before the interaction.
Summing the two quantities together one obtains:
\ba
\Sigma(i)=\Delta S_S(i)-\beta\,\delta Q(i),
\ea
where we have exploited the fact that the initial ancilla state $\hat{\eta}_{E_i}^{in}$ is thermal.
The second principle of thermodynamics can thus be expressed as:
\ba
\Sigma(i)\geq0\quad\forall i.
\ea
%The entropy production during a collision between the system $S$ and an ancilla is instead given by~\cite{strasberg}:
%\ba
%\Sigma=I(\hat{U}_{SE_{i+1}}\hat{R}_{S\cal E}(i)\hat{U}_{SE_{i+1}}^\dag)+S(\hat{\eta}_{E_{i+1}}^{out}||\hat{\eta}_{E_{i+1}})
%\ea
%where the mutual information is defined as:
%\ba
%&&I(\hat{U}_{SE_{i+1}}\hat{R}_{S\cal E}(i)\hat{U}_{SE_{i+1}}^\dag)=\\
%\nonumber
%&&S[\hat{\rho}_S(i+1)]+S(\hat{\eta}_{E_{i+1}}^{out})-S(\hat{U}_{SE_{i+1}}\hat{R}_{S\cal E}(i)\hat{U}_{SE_{i+1}}^\dag)
%\ea
%while the relative entropy reads:
%\ba
%S(\hat{\eta}_{E_{i+1}}^{out}||\hat{\eta}_{E_{i+1}})=-S(\hat{\eta}_{E_{i+1}}^{out})-\Tr\left[\hat{\eta}_{E_{i+1}}^{out}\log\hat{\eta}_{E_{i+1}}^{in}\right].\quad
%\ea

\subsection{Thermodynamics of the system in Example 2}

\begin{figure}[!t]
\centering
\includegraphics[width=\columnwidth]{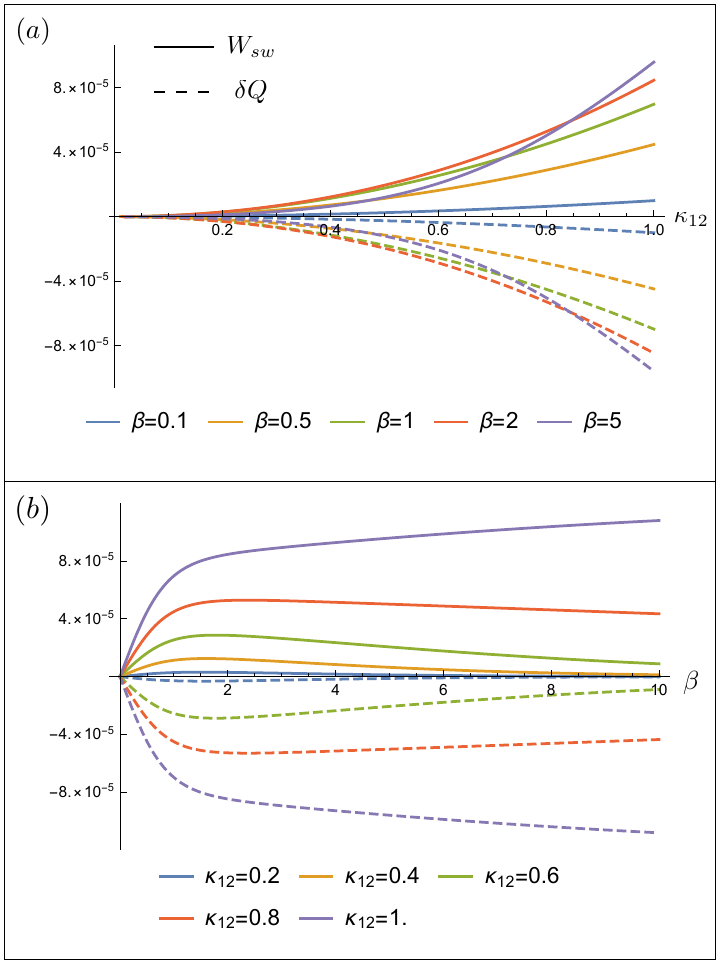}
\caption{Plot of $W_{sw}$ and $\delta Q$ at the steady state. In both plots $\omega_S=1$, $\omega_1=0.5$, $\omega_2=1.5$, $\alpha=0.2$. In panel (a) the quantities are plotted as a function of $\kappa_{12}$ for different values of the environmental temperature. It can be noticed that the work cost for having coherence in the steady state is always equal to the heat dissipated into the environment. Moreover, the amount of work is proportional to the amount of coherence created in system $S$. In panel (b) the same quantities are plotted as a function of the environmental temperature $\beta$ for different values of $\kappa_{12}$. Also in this case it can be noticed that the work cost is higher for lower temperatures, that is, for larger amounts of coherence.}
\label{fig:work_heat_steady}
\end{figure}

We now want to apply the formulas derived in the previous section to the system in Sec.~\ref{sec:example_2}. By substituting the Hamiltonians in Eqs.(~\ref{eq:interacting_qubits_hamiltonian},~\ref{eq:system_hamiltonian_ex_2},~\ref{eq:int_hamiltonian_ex_2}) we obtain explicit analytical expressions for the corresponding quantities, which are reported in App.~\ref{sec:appendix_3}.

The first term in the switching work is simply due to the mismatch of the energy difference $E_3-E_2$ in the structured ancilla and the system's energy separation $\omega_S$ and is common to all collision models. The last two terms arise because of the specific form of the interaction Hamiltonian of the system with the structured ancillae.

In Fig.~\ref{fig:thermo_plots} we plot $\Delta U$, $\delta Q$ and $W_{sw}$ for the initial part of the dynamics. One can observe, consistently with Eqs.~(~\ref{eq:int_en_var_example_2},~\ref{eq:switching_work_ex_2}), that since $\Delta U$ and $W_{sw}$ depend on the coherences in system $S$, both quantities presents oscillations before reaching their steady state values. Conversely, as the dissipated heat only depends on $\langle\hat{\sigma}_z\rangle$ and on the ancillae coherences, no oscillations are observed in this case. 
%Finally, the internal energy variation becomes zero after the system reaches equilibration, and thus also the switching work stops oscillating and simply becomes equal to the opposite of the dissipated heat.

After reaching the steady state, the variation of internal energy $\Delta U$ becomes identically zero. In the plots in Fig.~\ref{fig:work_heat_steady} one can observe that in all cases, at the steady state, a positive non-null amount of work is injected into the system in order to keep the steady state coherence. This energy is then exactly dissipated as heat into the thermal bath. This is perfectly consistent with the first principle of thermodynamics, which after the reaching of the steady state simply becomes $W_{sw}=-\delta Q$. This is also consistent with the second law of thermodynamics: no work can be created with a device operating with only one thermal bath. Moreover, in both plots one can observe that the amount of work is proportional to the amount of coherence created in the steady state, i.e. the larger the coherence the larger the work needed.

\begin{figure}[!t]
\centering
\includegraphics[width=\columnwidth]{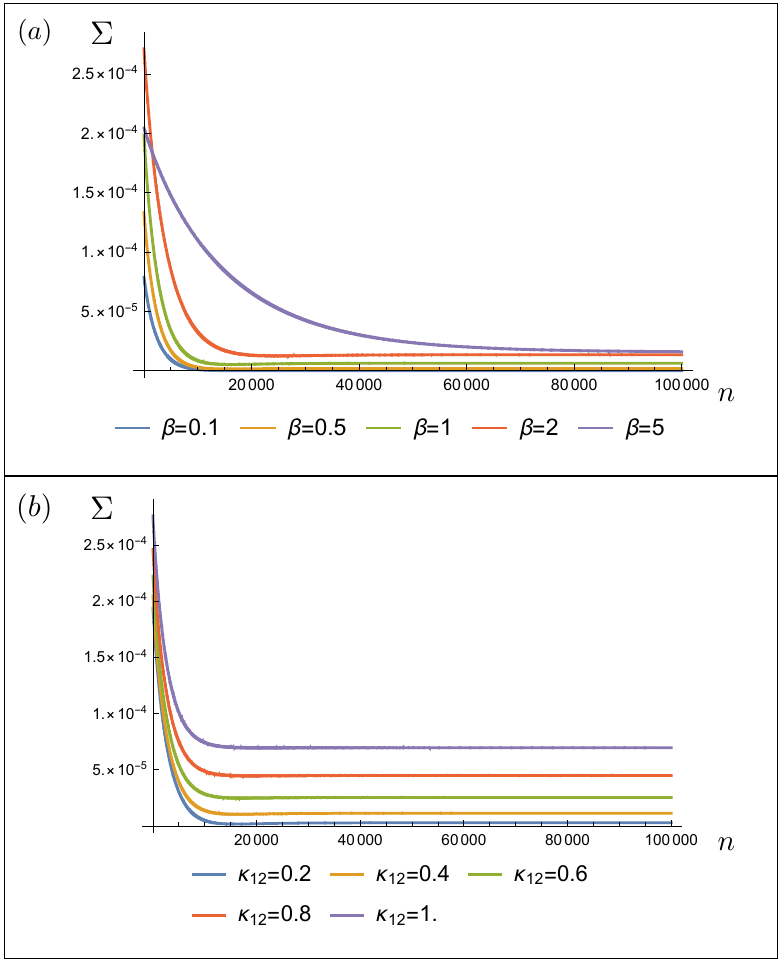}
\caption{Plot of the entropy production $\Sigma(i)$ at each step of the dynamics. Panel (a): plot of $\Sigma(i)$ for $\kappa_{12}=0.3$ for different values of $\beta$. One can notice that at the steady state (for large $n$) the entropy variation is larger for larger values of the coherence in the steady state. Panel (b): entropy  production plotted for $\beta=1$ and for different values of $\kappa_{12}$. Also in this case one can observe that the entropy increase is larger for larger values of the coherence in the steady state. In both cases one can also observe that the entropy variation is always positive, thus fulfilling the second principle of thermodynamics at every step of the dynamics.}
\label{fig:entropy_variation}
\end{figure}

Furthermore, in Fig.~\ref{fig:entropy_variation} we have plotted the step-by-step entropy production $\Sigma(i)$. It is immediate to note that this quantity is always positive, in fulfillment of the second principle of thermodynamics, and that it does never get to zero, since even at the steady state work has to be injected into the system and then dissipated as heat in order to retain coherence.
%%%%%%%%%%%%%%%%%%%%%%%%%%%%%%%%%%%%%%%%%%%%%%%%%%%%
\section{Conclusions\label{sec:conclusions}}
%%%%%%%%%%%%%%%%%%%%%%%%%%%%%%%%%%%%%%%%%%%%%%%%%%%%
In this work we have introduced a collision model in which the ancillae are composite systems, rather than simple TLS or harmonic oscillators. We have seen how to compute the discrete master equation for an open quantum system interacting with a structured environment using our model. We have shown how the presence of structured ancillae has consequences on the dynamics of $S$, most noticeably the possibility of creating coherence in the system in the steady state even if the environment is in a thermal state.
We have seen practically through a couple of examples involving qubits how the presence of the structured environment can lead to additional terms in the master equation, in particular the unitary driving terms leading to the steady state coherence, and how this is possible only when the system interacts with the degrees of freedom of the environment that possess coherence.

Furthermore, we have shown how the interaction between the system and the degrees of freedom of the environment possessing coherence requires a work cost to be activated, thus leading to the natural fulfillment of the first and second principle of thermodynamics without invoking any further assumption. We have also been able to compute analytically all the relevant thermodynamic quantities, and shown that one needs to keep injecting work into the system in order to retain the coherence in the steady state.

In future works we plan to use the model presented here to study the efficiency of heat engines and refrigerators exploiting coherence, and how including the cost of creating coherence in the system influences quantities such as the efficiency and the efficiency at maximum power.
%Moreover, we also plan to study the present model in the case of harmonic oscillators instead of TLS, and verify whether the presence of a structured environment has consequences on coherence and squeezing production and how to quantify their thermodynamic cost.
Finally, the present model might be useful also in different context than thermodynamics, for instance for investigating whether the presence of coherence in the environment can influence non Markovian dynamics or whether more complicated structured ancillae can be exploited to further engineer the steady state of the open quantum systems in order to obtain useful resources for quantum technologies.

\section*{Acknowledgements}
SC thanks D. Tamascelli for a useful comment during IQIS 2022 conference. SC acknowledges support by the Foundation for Polish Science (IRAP project, ICTQT, Contract No. 2018/MAB/5, cofinanced by the EU within the Smart Growth Operational Programme). G.D.C. acknowledges the support by the UK EPSRC EP/S02994X/1, the Royal Society IEC\textbackslash R2\textbackslash222003.

\onecolumngrid
\begin{appendix}
%%%%%%%%%%%%%%%%%%%%%%%%%%%%%%%%%%%%%%%%%%%%%%%%%
\section{Calculations of Example 1\label{app:example_1}}
%%%%%%%%%%%%%%%%%%%%%%%%%%%%%%%%%%%%%%%%%%%%%%%%%
In order to compute the constants $g_k^{(j)}$ and $g_{kk'}^{(j,j')}$ appearing in Eq.~\eqref{eq:2_qubits_ancillas_me} we define the non-interacting energy base of the 2 qubits as:
\ba
\nonumber
\ket{E_1}=\ket{\downarrow\downarrow},\;\ket{E_2}=\ket{\downarrow\uparrow},\\
\ket{E_3}=\ket{\uparrow\downarrow},\;\ket{E_4}=\ket{\uparrow\uparrow}.
\ea
The energy eigenbasis of the Hamiltonian $\hat{H}_{E_i}$ can be easily computed through standard methods, and is defined as:
\ba
\ket{E'_1}=\ket{E_1},\\
\ket{E'_2}=\left(a_-\ket{E_2}+b_-\ket{E_3}\right),\\
\ket{E'_3}=\left(a_+\ket{E_2}+b_+\ket{E_3}\right),\\
\ket{E'_4}=\ket{E_4},
\ea
where the constants $a_\pm,b_\pm$ are worth:
\ba
\label{eq:a_g_pm}
a_-=\frac{\kappa_{12}}{\sqrt{\kappa_{12}^2+(E'_2-y)^2}},\quad b_-=\frac{E'_2-y}{\sqrt{\kappa_{12}^2+(E'_2-y)^2}},\\
\nonumber
a_+=\frac{\kappa_{12}}{\sqrt{\kappa_{12}^2+(E'_3-y)^2}},\quad b_+=\frac{E'_3-y}{\sqrt{\kappa_{12}^2+(E'_3-y)^2}}.
\ea
The two energy basis are related via the unitary $\hat{V}$ as:
\ba
\ket{E'_k}=\sum_{i}V_{ik}\ket{E_i},
\ea
where the unitary $\hat{V}$ is worth:
\ba
\hat{V}=\begin{bmatrix}
1&0&0&0\\
0&b_-&b_+&0\\
0&a_-&a_+&0\\
0&0&0&1
\end{bmatrix}.
\ea
At this point we are ready to compute the constants $g_k^{(j,j')},g_{kk'}^{(j,j')}$. The expression for the thermal state is:
\ba
\hat{\eta}_{E_i}=\frac{e^{-\beta\hat{H}_{E_i}}}{Z}&=&\frac{1}{Z}\sum_ke^{-\beta E'_k}\dyad{E'_k}=\sum_{i,j,k}e^{-\beta E'_k}V_{ik}V_{kj}^\dag\dyad{E_i}{E_j}.
\ea
To illustrate how the calculation works, we show how to compute $g_1^{(+)}$:
\ba
\nonumber
g_1^{(+)}&=&\alpha_1\Tr\left[\hat{\sigma}_1^+\hat{\eta}_{E_i}\right]=\frac{\alpha_1}{Z}\Tr\left[\hat{\sigma}_1^+\sum_{i,j,k}e^{-\beta E'_k}V_{ik}V^\dag_{kj}\dyad{E_i}{E_j}\right]\\
\nonumber
&=&\frac{\alpha_1}{Z}\Tr\left[\sum_{j,k}e^{-\beta E'_k}\left(V_{1k}V^\dag_{kj}\dyad{E_3}{E_j}+V_{2k}V^\dag_{kj}\dyad{E_4}{E_j}\right)\right]\\
&=&\frac{\alpha_1}{Z}\sum_{k}e^{-\beta E'_k}\left(V_{1k}V^\dag_{k3}+V_{2k}V^\dag_{k4}\right)=0.
\ea
and similarly for $g_1^{(-)},g_2^{(\pm)}$.

As for $g_{11}^{(+,-)}$ we have:
\ba
\nonumber
&&g_{11}^{(+,-)}=\alpha_1^2\Tr\left[\hat{\sigma}_1^-\hat{\sigma}_1^+\hat{\eta}_{E_i}\right]=\frac{\alpha_1^2}{Z}\Tr\left[\hat{\sigma}_1^-\hat{\sigma}_1^+\sum_{i,j,k}e^{-\beta E'_k}V_{ik}V_{kj}^\dag\dyad{E_i}{E_j}\right]\\
\nonumber
&&=\frac{\alpha_1^2}{Z}\Tr\left[(\dyad{E_1}+\dyad{E_2})\sum_{i,j,k}e^{-\beta E'_k}V_{ik}V_{kj}^\dag\dyad{E_i}{E_j}\right]\\
\nonumber
&&=\frac{\alpha_1^2}{Z}\sum_ke^{-\beta E'_k}\left(V_{1k}V^\dag_{k1}+V_{2k}V^\dag_{k2}\right)\\
&&=\frac{\alpha_1^2}{Z}\left(e^{-\beta E'_1}+|b_-|^2e^{-\beta E'_2}+|b_+|^2e^{-\beta E'_3}\right).
\ea
and one can compute in a similar manner:
\ba
g_{11}^{(-,+)}&=&\alpha_1^2\Tr\left[\hat{\sigma}_1^+\hat{\sigma}_1^-\hat{\eta}_{E_i}\right]=\frac{\alpha_1^2}{Z}\left(|a_-|^2e^{-\beta E'_2}+|a_+|^2e^{-\beta E'_3}+e^{-\beta E'_4}\right),\\
g_{22}^{(+,-)}&=&\alpha_2^2\Tr\left[\hat{\sigma}_2^-\hat{\sigma}_2^+\hat{\eta}_{E_i}\right]=\frac{\alpha_2^2}{Z}\left(e^{-\beta E'_1}+|a_-|^2e^{-\beta E'_2}+|a_+|^2e^{-\beta E'_3}\right),\\
g_{22}^{(-,+)}&=&\alpha_2^2\Tr\left[\hat{\sigma}_2^+\hat{\sigma}_2^-\hat{\eta}_{E_i}\right]=\frac{\alpha_2^2}{Z}\left(|b_-|^2e^{-\beta E'_2}+|b_+|^2e^{-\beta E'_3}+e^{-\beta E'_4}\right).
\ea

Finally, we compute the cross term $g_{12}^{(+,-)}$ as:
\ba
\nonumber
g_{12}^{(+,-)}&=&g_{21}^{(+,-)*}=\alpha_1\alpha_2^*\Tr\left[\hat{\sigma}_1^-\hat{\sigma}_2^+\hat{\eta}_{E_i}\right]=\frac{\alpha_1\alpha_2}{Z}\Tr\left[\dyad{E_2}{E_3}\sum_{i,j,k}e^{-\beta E'_k}V_{ik}V_{kj}^\dag\dyad{E_i}{E_j}\right]\\
&=&\frac{\alpha_1\alpha_2}{Z}\sum_ke^{-\beta E'_k}V_{3k}V_{k2}^\dag=\frac{\alpha_1\alpha_2}{Z}\left(e^{-\beta E'_2}a_-b_-^*+e^{-\beta E'_3}a_+b_+^*\right).
\ea
Similarly we can compute the other cross terms:
\ba
g_{12}^{(-,+)}&=&g_{21}^{(-,+)*}=\alpha_1^*\alpha_2\Tr\left[\hat{\sigma}_1^+\hat{\sigma}_2^+\hat{\eta}_{E_i}\right]=\frac{\alpha_1\alpha_2}{Z}\left(e^{-\beta E'_2}b_-a_-^*+e^{-\beta E'_3}b_+a_+^*\right),\\
g_{12}^{(+,+)}&=&g_{21}^{(+,+)}=\alpha_1\alpha_2\Tr\left[\hat{\sigma}_1^-\hat{\sigma}_2^-\hat{\eta}_{E_i}\right]=0,\\
g_{12}^{(-,-)}&=&g_{21}^{(-,-)}=\alpha_1\alpha_2\Tr\left[\hat{\sigma}_1^+\hat{\sigma}_2^+\hat{\eta}_{E_i}\right]=0.
\ea

%%%%%%%%%%%%%%%%%%%%%%%%%%%%%%%%%%%%%%%%%%%%%
\section{Calculations of Example 2\label{sec:appendix_2}}
%%%%%%%%%%%%%%%%%%%%%%%%%%%%%%%%%%%%%%%%%%%%%
First we want to compute the factors $g_{E_i}^{(\pm)}$:
\ba
\nonumber
g_{E_i}^{(+)}&=&\alpha\Tr\left[\hat{\sigma}_{E_i}^+\hat{\eta}_{E_i}\right]=\frac{\alpha}{Z}\Tr\left[\dyad{E_3}{E_2}\sum_ke^{-\beta E'_k}\dyad{E'_k}\right]=\frac{\alpha}{Z}\Tr\left[\dyad{E_3}{E_2}\sum_{i,j,k}V_{ik}V^\dag_{kj}e^{-\beta E_i}\dyad{E_j}\right]\\
&=&\frac{\alpha}{Z}\sum_{k}e^{-\beta E'_k}V_{2k}V^\dag_{k3}=\frac{\alpha}{Z}\left[e^{-\beta E'_2}a_-b_-^*+e^{-\beta E'_3}a_+b_+^*\right],
\ea
and similarly we get:
\ba
g_{E_i}^{(-)}=\frac{\alpha^*}{Z}\left[e^{-\beta E'_2}a_-^*b_-+e^{-\beta E'_3}a_+^*b_+\right]=g_{E_i}^{(+)*}.
\ea
Then we move to the calculation of the factors $g_{E_i}^{(\pm,\pm)}$, starting from $g_{E_i}^{(-,+)}$:
\ba
\nonumber
&&g_{E_i}^{(-,+)}=\Tr\left[\hat{\sigma}_{E_i}^+\hat{\sigma}_{E_i}^-\hat{\eta}_{E_i}\right]=\frac{|\alpha|^2}{Z}\Tr\left[\dyad{E_3}{E_2}\dyad{E_2}{E_3}\sum_ke^{-\beta E'_k}\dyad{E'_k}\right]\\
&=&\frac{|\alpha|^2}{Z}\Tr\left[\dyad{E_3}{E_3}\sum_{i,j,k}e^{-\beta E'_k}V_{ik}V^\dag_{kj}\dyad{E_i}{E_j}\right]=\frac{|\alpha|^2}{Z}\sum_{k}e^{-\beta E'_k}V_{3k}V^\dag_{k3}=\frac{|\alpha|^2}{Z}\left[e^{-\beta E'_2}|a_-|^2+e^{-\beta E'_3}|a_+|^2\right],
\ea
and similarly for $g_{E_i}^{(+,-)}$ we find:
\ba
g_{E_i}^{(+,-)}=\frac{|\alpha|^2}{Z}\left[e^{-\beta E'_2}|b_-|^2+e^{-\beta E'_3}|b_+|^2\right].
\ea
Finally, it is immediate to see that:
\ba
g_{E_i}^{(+,+)}=\Tr\left[\hat{\sigma}_{E_i}^+\hat{\sigma}_{E_i}^+\hat{\eta}_{E_i}\right]=0,\\
g_{E_i}^{(-,-)}=\Tr\left[\hat{\sigma}_{E_i}^-\hat{\sigma}_{E_i}^-\hat{\eta}_{E_i}\right]=0.
\ea

The matrix elements are mapped by the master equation in~\eqref{eq:master_equation_2} according to:
\ba
&&\rho_{11}\rightarrow\rho_{11}+\delta t^2\left(g_{E_i}^{(-,+)}\rho_{22}-g_{E_i}^{(+,-)}\rho_{11}\right)+\frac{\delta t^2}{2}\omega_S\left(g_{E_i}^{(+)}\rho_{12}+g_{E_i}^{(-)}\rho_{21}\right)+i\delta t\left(g_{E_i}^{(+)}\rho_{12}-g_{E_i}^{(-)}\rho_{21}\right),\\
&&\rho_{22}\rightarrow\rho_{22}-\delta t^2\left(g_{E_i}^{(-,+)}\rho_{22}-g_{E_i}^{(+,-)}\rho_{11}\right)-\frac{\delta t^2}{2}\omega_S\left(g_{E_i}^{(+)}\rho_{12}+g_{E_i}^{(-)}\rho_{21}\right)-i\delta t\left(g_{E_i}^{(+)}\rho_{12}-g_{E_i}^{(-)}\rho_{21}\right),\\
&&\rho_{12}\rightarrow\rho_{12}+\frac{\delta t^2}{2}\omega_Sg_{E_i}^{(-)}(\rho_{11}-\rho_{22})+i\delta tg_{E_i}^{(-)}(\rho_{11}-\rho_{22})-i\delta t\omega_S\rho_{12}-\frac{\delta t^2}{2}\rho_{12}\left(g_{E_i}^{(+,-)}+g_{E_i}^{(-,+)}+\omega_S^2\right),\qquad\\
&&\rho_{21}\rightarrow\rho_{21}+\frac{\delta t^2}{2}\omega_Sg_{E_i}^{(+)}(\rho_{11}-\rho_{22})-i\delta tg_{E_i}^{(+)}(\rho_{11}-\rho_{22})+i\delta t\omega_S\rho_{21}-\frac{\delta t^2}{2}\rho_{21}\left(g_{E_i}^{(+,-)}+g_{E_i}^{(-,+)}+\omega_S^2\right).
\ea

The steady state values of $\langle\hat{\sigma}_x\rangle$, $\langle\hat{\sigma}_y\rangle$, $\langle\hat{\sigma}_z\rangle$ are:
\ba
%\langle\hat{\sigma}_S^z\rangle&=&\frac{(g\delta t)^2\left(g_{E_i}^{(-,+)}+g_{E_i}^{(+,-)}\right)\left(g_{E_i}^{(-,+)}-g_{E_i}^{(+,-)}\right)}{K}\qquad\\
%\langle\hat{\sigma}_x\rangle&=&-\frac{2i(g\delta t)\left(g_{E_i}^{(-)}-g_{E_i}^{(+)}\right)\left(g_{E_i}^{(-,+)}-g_{E_i}^{(+,-)}\right)}{K}\\
%\langle\hat{\sigma}_y\rangle&=&-\frac{2(g\delta t)\left(g_{E_i}^{(+)}+g_{E_i}^{(-)}\right)\left(g_{E_i}^{(-,+)}-g_{E_i}^{(+,-)}\right)}{K}\\
%K&=&(g\delta t)^2\left(g_{E_i}^{(+,-)}+g_{E_i}^{(-,+)}\right)^2+8g_{E_i}^{(+)}g_{E_i}^{(-)}
&&\langle\hat{\sigma}_z\rangle=\frac{\left(g_E^{(-,+)}-g_E^{(+,-)}\right)\left[4\omega_S^2+\delta t^2\left(g_E^{(-,+)}+g_E^{(+,-)}\omega_S^2\right)^2\right]}{K},\\
&&\langle\hat{\sigma}_x\rangle=\frac{\left(g_E^{(-,+)}-g_E^{(+,-)}\right)2\Re\left[g_E^{(-)}\left(4\omega_S+2i\delta t\left(g_{E}^{(-,+)}+g_E^{(+,-)}\right)+\delta t^2\omega_S\left(g_{E}^{(-,+)}+g_E^{(+,-)}+\omega_S^2\right)\right)\right]}{K},\\
&&\langle\hat{\sigma}_y\rangle=-\frac{\left(g_E^{(-,+)}-g_E^{(+,-)}\right)2\Im\left[g_E^{(-)}\left(4\omega_S+2i\delta t\left(g_{E}^{(-,+)}+g_E^{(+,-)}\right)+\delta t^2\omega_S\left(g_{E}^{(-,+)}+g_E^{(+,-)}+\omega_S^2\right)\right)\right]}{K},\qquad\quad\\
\nonumber
&&\quad K=8g_E^{(-)}g_E^{(+)}\left(g_E^{(-,+)}+g_E^{(-,+)}-\omega_S^2\right)+4\omega_S^2\left(g_E^{(-,+)}+g_E^{(+,-)}\right)\\
&&\quad+\delta t^2\left(g_E^{(-,+)}+g_E^{(+,-)}\right)\left[\left(g_E^{(-,+)}+g_E^{(+,-)}\right)^2+2\omega_S^2\left(g_E^{(-,+)}+g_E^{(+,-)}-g_E^{(-)}g_E^{(+)}\right)+\omega_S^4\left(1-2g_E^{(-)}g_E^{(+)}\right)\right].
\ea

\section{Thermodynamic expressions\label{sec:appendix_3}}

Explicit expressions for Eqs.(~\ref{eq:int_en_var_explicit},~\ref{eq:heat_explicit}):
\ba
\nonumber
&&\Delta U=-i\delta t\sum_{j,k}g_k^{(j)}\Tr\left[\comm{\hat{H}_S}{\hat{A}_S^{(k,j)}}\hat{\rho}_S(i)\right]\frac{\delta t^2}{2}\sum_{k,j}g_k^{(j)}\Tr\left[\comm{\hat{A}_S^{(k,j)\dag}}{\hat{H}_S}\hat{H}_S\hat{\rho}_S(i)\right]\\
\nonumber
&&+\frac{\delta t^2}{2}\sum_{k,j}g_k^{(j)}\Tr\left[\comm{\hat{H}_S}{\hat{A}_S^{(k,j)}}\hat{\rho}_S(i)\hat{H}_S\right]+\frac{g\delta t^2}{2}\sum_{k,k',j,j'}g_{kk'}^{(j,j')}\Tr\left[\left(2\hat{A}_S^{(k,j)}\hat{H}_S\hat{A}_S^{(k,j)\dag}-\acomm{\hat{A}_S^{(k,j)\dag}\hat{A}_S^{(k,j)}}{\hat{H}_S}\right)\hat{\rho}_S(i)\right]\\
&&+\frac{\delta t^2}{2}\sum_{j,k}\Tr\left[\comm{\hat{H}_S}{g_{kH_E}^{(j)}\hat{A}_S^{(k,j)}-g_{kH_E}^{(j)*}\hat{A}_S^{(k,j)\dag}}\hat{\rho}_S(i)\right],
\ea
for the variation of internal energy, while the heat exchanged reads:
\ba
\nonumber
&&\delta Q=i\delta t\sum_{k,j}(g_{kH_E}^{(j)}-g_{kH_E}^{(j)*})\Tr\left[\hat{A}_S^{(k,j)}\hat{\rho}_S(i)\right]-\frac{\delta t^2}{2}\sum_{j,k}\Tr\left[\comm{\hat{H}_S}{g_{kH_E}^{(j)}\hat{A}_S^{(k,j)}-g_{kH_E}^{(j)*}\hat{A}_S^{(k,j)\dag}}\hat{\rho}_S(i)\right]\\
\nonumber
&&+\frac{\delta t^2}{2}\sum_{j,k}\Tr\left[\hat{A}_S^{(j,k)}\hat{\rho}_S(i)\right]\Tr\left[(2\hat{H}_E\hat{B}_{Ek}^{(j)}\hat{H}_E-\acomm{\hat{H}_E^2}{\hat{B}_{Ek}^{(j)}})\hat{\eta}_{E_{i+1}}\right]\\
\nonumber
&&-\frac{\delta t^2}{2}\sum_{k,k',j,j'}\Tr\left[\hat{A}_S^{(k,j)\dag}\hat{A}_S^{(k,j)}\hat{\rho}_S\right]\Tr\left[\comm{\hat{H}_E}{\hat{B}_{Ek}^{(j)}}\hat{\eta}_{E_{i+1}}\hat{B}_{Ek}^{(j)\dag}\right]+\frac{\delta t^2}{2}\sum_{k,k',j,j'}\Tr\left[\hat{A}_S^{(k,j)}\hat{A}_S^{(k,j)\dag}\hat{\rho}_S\right]\Tr\left[\comm{\hat{H}_E}{\hat{B}_{Ek}^{(j)}}\hat{B}_{Ek}^{(j)\dag}\hat{\eta}_{E_{i+1}}\right].\\
\ea
Finally, we will not write explicitly the switching work in Eq.~\eqref{eq:switching_work_explicit}, since one can easily derive it from:
\ba
W_{sw}=\Delta U-\delta Q.
\ea

The explicit expressions of $\Delta U$, $\delta Q$ and $W_{sw}$ for the system in Example 2 read:
\ba
\label{eq:int_en_var_example_2}
\Delta U=&-&\frac{\delta t^2}{2}\omega_S\left[\expval{\hat{\sigma}_S^z}\left(g_{E_i}^{(+,-)}+g_{E_i}^{(-,+)}\right)+\left(g_{E_i}^{(+,-)}-g_{E_i}^{(-,+)}\right)\right]\\
\nonumber
&+&\frac{\delta t^2}{2}\omega_S^2\left(g_{E_i}^{(+)}\rho_{12}+g_{E_i}^{(-)}\rho_{21}\right)+i\delta t\omega_S\left(g_{E_i}^{(+)}\rho_{12}-g_{E_i}^{(-)}\rho_{21}\right),\\
\label{eq:heat_example_2}
\delta Q&=&-\frac{\delta t^2}{2}(E_3-E_2)\left[\expval{\hat{\sigma}_S^z}\left(g_{E_i}^{(+,-)}+g_{E_i}^{(-,+)}\right)+\left(g_{E_i}^{(+,-)}-g_{E_i}^{(-,+)}\right)\right]+\frac{\delta t^2}{2}\kappa_{12}|\alpha|^2\left(\eta_{23}+\eta_{32}\right),\\
\label{eq:switching_work_ex_2}
W_{sw}=&&\frac{\delta t^2}{2}(E_3-E_2-\omega_S)\left[\expval{\hat{\sigma}_S^z}\left(g_{E_i}^{(+,-)}+g_{E_i}^{(-,+)}\right)+\left(g_{E_i}^{(+,-)}-g_{E_i}^{(-,+)}\right)\right]\\
\nonumber
&-&\frac{\delta t^2}{2}\kappa_{12}|\alpha|^2\left(\eta_{23}+\eta_{32}\right)+\frac{\delta t^2}{2}\omega_S^2\left(g_{E_i}^{(+)}\rho_{12}+g_{E_i}^{(-)}\rho_{21}\right)+i\delta t\omega_St\left(g_{E_i}^{(+)}\rho_{12}-g_{E_i}^{(-)}\rho_{21}\right).
\ea

\end{appendix}
\end{document}